\documentclass[fleqn,usenatbib]{mnras}

\usepackage{newtxtext,newtxmath}

\usepackage[T1]{fontenc}

\DeclareRobustCommand{\VAN}[3]{#2}
\let\VANthebibliography\thebibliography
\def\thebibliography{\DeclareRobustCommand{\VAN}[3]{##3}\VANthebibliography}

\usepackage{graphicx}	
\usepackage{amsmath}	
\usepackage{graphicx}
\usepackage{subcaption}
\usepackage{gensymb}
\usepackage{xtab,afterpage}
\usepackage{lastpage}
\usepackage{longtable}

\title[Distances to PHANGS Galaxies]{Distances to PHANGS Galaxies: New Tip of the Red Giant Branch Measurements and Adopted Distances}

\author[G. S. Anand et al.]{
Gagandeep S. Anand$^{1,2}$\thanks{E-mail: gsanand@hawaii.edu}\thanks{IPAC Visiting Graduate Student Fellow},
Janice C. Lee$^{1}$,
Schuyler D. Van Dyk$^{1}$,
Adam K. Leroy$^{3}$,
\newauthor
Erik Rosolowsky$^{4}$,
Eva Schinnerer$^{5}$,
Kirsten Larson$^{1}$,
Ehsan Kourkchi$^{2}$,
\newauthor
Kathryn Kreckel$^{6}$,
Fabian Scheuermann$^{6}$,
Luca Rizzi$^{7}$,
David Thilker$^{8}$,
R. Brent Tully$^{2}$,
\newauthor
Frank Bigiel$^{9}$,
Guillermo A. Blanc$^{10,11}$,
Médéric Boquien$^{12}$,
Rupali Chandar$^{13}$,
\newauthor
Daniel Dale$^{14}$,
Eric Emsellem$^{15,16}$,
Sinan Deger$^{1}$,
Simon C.O. Glover$^{17}$,
Kathryn Grasha$^{18}$,
\newauthor
Brent Groves$^{18,19}$,
Ralf S.\ Klessen$^{17,20}$,
J.~M.~Diederik Kruijssen$^{21}$,
Miguel Querejeta$^{22}$,
\newauthor
Patricia S\'anchez-Bl\'azquez$^{23}$,
Andreas Schruba$^{24}$,
Jordan Turner$^{14}$,
Leonardo Ubeda$^{25}$,
\newauthor
Thomas G. Williams$^{5}$,
Brad Whitmore$^{25}$
\\
Affiliations are listed at the end of the paper.}

\date{\today}

\pubyear{2020}

\begin{document}
\label{firstpage}
\pagerange{\pageref{firstpage}--\pageref{lastpage}}
\maketitle

\begin{abstract}
PHANGS-HST is an ultraviolet-optical imaging survey of 38 spiral galaxies within $\sim$20 Mpc. Combined with the PHANGS-ALMA, PHANGS-MUSE surveys and other multiwavelength data, the dataset will provide an unprecedented look into the connections between young stars, H~{\sc ii} regions, and cold molecular gas in these nearby star-forming galaxies. Accurate distances are needed to transform measured observables into physical parameters (e.g., brightness to luminosity, angular to physical sizes of molecular clouds, star clusters and associations). PHANGS-HST has obtained parallel ACS imaging of the galaxy halos in the F606W and F814W bands. Where possible, we use these parallel fields to derive tip of the red giant branch (TRGB) distances to these galaxies. In this paper, we present TRGB distances for 11 galaxies from $\sim$4 to $\sim$15 Mpc, based on the first year of PHANGS-HST observations. Five of these represent the first published TRGB distance measurements (IC~5332, NGC~2835, NGC~4298, NGC~4321, and NGC~4328), and eight of which are the best available distances to these targets. We also provide a compilation of distances for the 118 galaxies in the full PHANGS sample, which have been adopted for the first PHANGS-ALMA public data release. \end{abstract}





\section{Introduction}
The observed velocity of a galaxy consists of two components. The first is its recessional velocity due to the expansion of the universe, i.e., the Hubble flow. This portion of the observed velocity is simply given by the Hubble constant times its distance ($\mathrm{H_{0}}D$). The second component of the observed velocity is due to gravitational interactions with other objects, which is referred to as a galaxy's peculiar velocity ($v_{\mathrm{pec}}$). Taken together, the observed velocity of a galaxy is given by
\begin{equation}
    v_{\mathrm{obs}} = H_{0}D + v_{\mathrm{pec}}.
\end{equation}

In the nearby universe (i.e., within a few tens of Mpc), the peculiar velocity can be a substantial component of the observed velocity. This means distances based solely on the recessional velocity are subject to large systematic errors. This issue necessitates the use of redshift-independent distances, such as those based on standard candles and rulers, for study of galaxies in the nearby Universe. 

In this paper, we present a curated set of redshift-independent distances for galaxies in the PHANGS\footnote{\url{www.phangs.org}} (Physics at High Angular Resolution in Nearby Galaxies) sample (A.~K.\ Leroy et al, in preparation). The distances presented in this paper are a combination of new tip of the red giant branch (TRGB) measurements based on \textit{Hubble Space Telescope} (\textit{HST}) imaging obtained by the PHANGS-HST survey (J.C.~Lee et al., in preparation) in its first year of observations (which began in 2019 April) and best available distances compiled from the literature.  

The goal of the PHANGS programme is to elucidate the physics that control the multi-scale process of star formation in galaxies.  The effort is built around PHANGS-ALMA (P.I. E. Schinnerer; A.~K.\ Leroy et al. in preparation), an ALMA survey of $N$=74 galaxies that includes all southern ($-75\degree$ < Dec < $20\degree$), face-on, massive, star-forming galaxies at distances (< 20~Mpc) where ALMA can resolve the molecular interstellar medium into individual molecular clouds (50--150~pc). This sample was observed via a Cycle 5 Large Programme (PI: E Schinnerer), and several smaller programs in Cycles 2-6. Extensions of the programme to additional galaxies that expand the covered parameter space are ongoing, and bring the current sample observed by ALMA to 89.  In addition, the broader PHANGS collaboration has studied a number of other nearby targets, including some northern (beyond the reach of ALMA) and edge-on targets. In total, PHANGS currently includes 118 targets of interest, and it is for this greater sample of galaxies that we provide ``best-available'' distances adopted for PHANGS analysis in this paper.

Distances to the PHANGS galaxies are essential to the main science goals of the PHANGS collaboration.  Nearly every derived parameter depends on the adopted distance, and robust distances are required for the basic transformation of angular size and brightness into physical sizes and absolute luminosities. Inaccurate distances will bias other quantities of interest, including star cluster and molecular cloud mass functions, luminosity functions, and dynamical mass-to-light ratios. Distances are also needed as inputs for producing ALMA products at specified physical (e.g., 60~pc, 150~pc) resolutions, which are vital for consistent galaxy-to-galaxy comparisons. 

In this paper, we use parallel imaging from PHANGS-HST to derive TRGB estimates to 11 galaxies, and also provide a careful literature compilation of best distances for the full PHANGS sample of interest (N=118). Prior to this work, accurate distances, based on standard candles, were available for $\sim$45\% of the full PHANGS sample of 118 galaxies. To this we add the first TRGB distance measurements for 5 galaxies, and additional TRGB measurements for 6 galaxies based on our \textit{HST} parallel observations. Eight of these new TRGB measurements represent the best available distances for these targets. In Section 2 we describe the parallel PHANGS-HST observations used in this work. In Section 3, we describe the TRGB  methodology for measuring distances from our parallel imaging, and then present our results. We present our selection of literature distances in Section 4, and end with a brief summary and future outlook in Section 5.

\section{PHANGS-HST Observations}
A subset of the overall PHANGS sample best suited for joint \textit{HST}-ALMA studies of resolved young stellar populations and clouds ($N$=38) were targeted by the Cycle 26 PHANGS-HST Treasury Programme (PI: J.C.~Lee, programme GO-15654). The PHANGS-HST sample was selected to be
\begin{itemize}
    \item Relatively face-on ($i<70\degree$), to minimise source blending and projected dust attenuation.
    \item Avoid the Galactic plane ($|b|\gtrsim15\degree$), to minimise the effects of Milky Way reddening and foreground stars.
    \item Have sufficient star formation activity(star formation rates $> 0.3 M_{\odot}/\mathrm{yr}$) to ensure widespread CO detections for joint analysis of clusters/associations and clouds.
    \item Nearby ($D$ $\lesssim$17~Mpc) to ensure high levels of spatial resolution (though in this paper we find several of the PHANGS-HST targets to likely lie beyond this initial criterion).
\end{itemize}
The combination of ALMA observations with those from \textit{HST} and MUSE (E.~Emsellem et al., in preparation) allows PHANGS to chart, for the first time, the connections between cold (molecular) gas and young stars on the fundamental scales of molecular clouds, young star clusters, and H~{\sc ii} regions, over a broad range of galactic environments in the nearby Universe \citep{2018ApJ...860..172S,2018ApJ...861L..18U,2019ApJ...887...49S,2019ApJ...887...80K,2020MNRAS.493.2872C}.

The primary goal of the PHANGS-HST observations (to be fully described in J.C.~Lee, in preparation) was to obtain UV-optical imaging of the resolved stellar populations within the star-forming disk. However, the observations also provide an incidental opportunity to observe the galaxy halo with the Advanced Camera for Surveys Wide Field Channel (ACS/WFC) in ``parallel" mode.  Our observations were designed so that ACS imaging in the F606W (a ``wide" $V$-band) and F814W (approximately $I$-band) filters accompanies each corresponding PHANGS-HST ``primary" observation with the Wide Field Camera 3 Ultraviolet/Visible (WFC3/UVIS) channel. 

For the range of distances and angular sizes of the spiral galaxies in the PHANGS-HST sample, the ACS field-of-view generally falls on the halo of the target galaxy when WFC3 is centered on the galaxy itself, though there is a range of potential outcomes. For galaxies with relatively large angular sizes, the parallel observations may include portions of the outer disk.  For galaxies with much smaller sizes, the parallels may lie too far to detect any sizeable halo population, which limits the usefulness of the field for TRGB analysis. Given that the science requirements of PHANGS-HST constrain the placement of the primary pointings, optimizing placement of the parallel fields, as would be pursued by a focused TRGB programme, is secondary, and is limited by the fixed focal plane and spatial offset of the two cameras of \textit{HST}.  

Figures \ref{fig:fp1} and \ref{fig:fp2} illustrate the positioning of the ACS field-of-view on each galaxy. Orientation (ORIENT) constraints were imposed, when possible, to prevent the parallel observation from sampling the galaxy disk, nearby galaxy neighbours, and/or extremely bright foreground stars. For some targets with large angular sizes for which it would be impossible to entirely avoid a large disk, we placed the parallels along the major axis to aid in disentangling the disk and halo. In several cases, such orient constraints needed to be lifted or considerably relaxed to enable enough guide stars to be acquired with the Fine Guidance Sensors. 

The 5-band primary observations with WFC3/UVIS were sequenced in each orbit to optimize exposure time in the parallel observations without impacting the primary observations. Each pointing of the telescope spanned 2 or 3 orbits, depending on whether archival observations of the inner galaxy (targeted by ALMA) were available from prior \textit{HST} programmes which matched the PHANGS-HST science requirements. For pointings spanning a two-orbit duration, the total exposure times in the ACS parallel V and I images are about 2100~s each, whilst for the three-orbit pointings they are about 3500~s and 3200~s, respectively. Exposure times for each pointing are provided in Table \ref{tab:exposure-times}. 

PHANGS-HST observations began in 2019 April and are expected to continue until 2021 May.  The analyses presented in this paper are based on the first year of observations through 2020 July, and include 37 pointings in 30 galaxies.  We will present the TRGB analysis based on parallel observations of the remaining 7 pointings (in six galaxies) in a future paper. All of these remaining pointings are observations which have been executed but failed, due to guide star acquisition issues, and are scheduled to be re-observed.



\section{Tip of the Red Giant Branch Measurements}

Low-mass stars ($<$2 $\mathrm{M_{\odot}})$ ascending the red giant branch (RGB) eventually reach a state when helium begins to fuse in the degenerate core via the triple-$\alpha$ process. At the end of this runaway process, the star rearranges itself, becomes less luminous, and appears on the horizontal branch. The maximum degenerate core mass is a constant, resulting in stars at the TRGB sharing the same maximum luminosity, modulo a colour-dependent term. This colour-dependence is largely the result of the effects of line-blanketing (dependent on metallicity), and to a lesser extent, age. In the best case scenarios, distances can be measured with the TRGB to accuracies of $\sim$5$\%$.

The standardizable candle nature of the TRGB has made it a powerful tool for determining extragalactic distances \citep{1993ApJ...417..553L,1995AJ....109.1645M,2018SSRv..214..113B}, and its popularity in the literature has been steadily increasing \citep{2016ApJ...827...89T,2017AJ....154...51M,2018ApJ...858...62K,2018ApJ...861L...6A, 2020ApJ...895L...4D}. At present, the CMDs/TRGB catalog on the Extragalactic Distance Database\footnote{\url{edd.ifa.hawaii.edu}} (EDD) hosts colour-magnitude diagrams and TRGB distances to nearly 500 galaxies \citep{2009AJ....138..332J}.

\subsection{Methodology}
There are two main techniques used in the literature to determine the location of the TRGB in a colour-magnitude diagram (CMD). Both involve constructing a luminosity function for stars above and below the TRGB, namely asymptotic giant branch (AGB) and RGB stars, respectively. The first popular method \citep{2000AJ....119.1197S, 2014ApJ...795L..35C, 2018A&A...615A..96M, 2019ApJ...875..136V} uses an edge detection algorithm (often a Sobel filter) to highlight the point of greatest discontinuity, which corresponds to the sharp change in the luminosity function occurring at the TRGB. The luminosity function may first be smoothed to suppress false edges arising from noise \citep{2019ApJ...885..141B}.

The second method commonly found in the literature involves fitting the luminosity function of the RGB and AGB population, typically with a broken-power law \citep{2002AJ....124..213M,2006AJ....132.2729M,2016ApJ...826...21M}. This method allows for the straightforward incorporation of results from artificial star experiments to account for photometric errors, incompleteness, and bias. Due to this benefit, we use this latter technique for our analysis, with the specific methodology described in detail by \cite{2006AJ....132.2729M}, and with updates provided by \cite{2014AJ....148....7W}. This overall procedure is the same as the one previously described in \cite{2009AJ....138..332J}. In the rest of this subsection, we briefly summarise our methodology.

We obtain the individual charge transfer efficiency (CTE)-corrected \textit{*.flc} frames from the Mikulski Archive for Space Telescopes\footnote{\url{https://mast.stsci.edu/}}. We perform PSF photometry on these individual exposures with DOLPHOT \citep{2000PASP..112.1383D,2016ascl.soft08013D}, which uses Tiny Tim PSFs \citep{1993ASPC...52..536K} and includes aperture corrections based on measurements of bright, isolated stars in each frame. We use the drizzled F814W (\textit{*.drc}) image as the reference frame for the alignment of the individual (\textit{*.flc }) images. In some instances the relative astrometry between individual frames is not good enough for DOLPHOT to obtain successful alignments between all the frames. For these cases, we first run the images through STScI's \textit{tweakreg} package until a satisfactory alignment (typical root-mean-square, rms, uncertainty of $\sim 0{\farcs}01$) is reached. Note that we do not pay attention to \textit{absolute} astrometry, but only \textit{relative} astrometry.

DOLPHOT outputs photometry for each individual exposure, as well as a set of photometry for the combination of individual \textit{*.flc} exposures. For our work, we use this combined photometry after applying a series of quality cuts for parameters including signal-to-noise ratio (SNR), crowding, and sharpness. For this paper, we use quality cuts adopted from \cite{2017AJ....154...51M}, except we increase the baseline total SNR cutoff in F606W from 2 to 5. For a few of our more distant targets (e.g., NGC~4321), we lower the SNR cutoff in F606W from 5 back to 2, and the F814W cutoff from 5 to 4, in order to increase the depth of the CMD below the TRGB. The specific crowding cuts adopted select for stars with $(Crowd_{\mathrm{F606W}}+Crowd_{\mathrm{F814W}}) < 0.8$, and for sharpness with $(Sharp_{\mathrm{F606W}}+Sharp_{\mathrm{F814W}})^{2} < 0.075$.

We also use DOLPHOT to perform artificial star experiments for each of the target fields. For each field, we insert and attempt to recover 100,000 artificial stars spanning the full range of magnitudes and colours seen in the measured photometry. These results allow us to quantify the true levels of error, completeness, and photometric bias present in the observed photometry. This is especially important, as it has been shown that DOLPHOT systematically underestimates its reported errors \citep{2014ApJS..215....9W}.

We proceed to fit a broken-power law to the luminosity function of the AGB and RGB populations, with the break denoting the location of the TRGB. The physical basis for this parameterization is the abrupt change in the observed luminosity function brought upon by stars undergoing the helium flash once they reach the TRGB. The results of the artificial stars are incorporated here by convolving the luminosity function with the completeness, error, and bias, as described in detail by \cite{2006AJ....132.2729M} and \cite{2014AJ....148....7W}. For some galaxies, we use the blue upper-main sequence stars as a proxy for all young stars to remove parts of the field before we perform our analysis. This is to reduce contamination from regions with large amounts of Population I stars, including red supergiants whose sharp feature lies on the blue edge of the RGB (see the galaxies presented in \cite{2019ApJ...872L...4A,2019ApJ...880...52A} for detailed examples). To further reduce contaminant stars in our sample, we limit the F606W$-$F814W of stars used in our fits-- these ranges are shown in Figures \ref{fig:ngc4826}--\ref{fig:newCMD} as the break between the red horizontal lines.

With the observed quantities in hand, we turn to the calibration for the TRGB obtained by \cite{2007ApJ...661..815R}. In addition to a zero-point TRGB calibration (anchored to a geometric calibration of the horizontal branch provided by \cite{2000ApJ...533..215C}), they provide a colour calibration for both WFPC2 and ACS flight filter systems for \textit{HST}. Combined together and in our choice of filters, these take the form of:
\begin{equation}
M_{ACS}^{F814W} = -4.06+0.20[(F606W-F814W)-1.23].
\end{equation}
We calculate the (F606W-F814W) term by taking the median color of stars within 0.05 mag below the measured TRGB, with the associated uncertainty determined via 1000 bootstrap resampling trials (as laid out in \citealt{2014AJ....148....7W}). Before applying this calibration, the observed magnitude and colour of the TRGB are corrected for foreground extinction \citep{2011ApJ...737..103S}. We do not account for any potential reddening intrinsic to the halos of these galaxies, though we note that previous studies \citep{2010MNRAS.405.1025M} have shown this effect to be quite small ($A_{I}\sim$0.01 mag, \citealt{2019ApJ...886...61Y}). Data taken in additional filters (e.g. near-infrared passbands with WFC3) could be used to determine the exact reddening to the TRGB stars themselves, as recently outlined by \cite{2020AJ....160..170M}.

The effects of line blanketing are minimised in the F814W filter for RGB stars, hence its use here and elsewhere in the literature. Over typical F606W$-$F814W colours seen in galaxy halos, the absolute magnitude of the TRGB varies by $\leq$0.1 mag. In bluer or redder bands, the effects can change the absolute magnitude by over one magnitude, increasing the dependence on the quality of the underlying calibration, and heightening the potential of systematic errors. The final errors on our reported distances combine in quadrature the statistical uncertainties in the measured quantities (including the dust maps, \citealt{2011ApJ...737..103S}) with an adopted 0.07~mag systematic uncertainty \citep{2007ApJ...661..815R, 2017AJ....154...51M} in the underlying absolute calibration. We note that there are several other calibrations available for the absolute magnitude of the TRGB \citep{2017ApJ...835...28J, 2019ApJ...882...34F, 2019ApJ...886...61Y, 2020arXiv200804181J}, which differ from our adopted calibration (and from each other) at the 0-5$\%$ level (depending on the underlying metallicity and age of the RGB). At present, the source of the disagreement is under debate. We adopt the \cite{2007ApJ...661..815R} calibration to retain consistency with the Extragalactic Distance Database, and reserve further discussion for future planned work on the matter.

The underlying photometry and complete list of derived parameters for the galaxies presented here are publicly available under the CMDs/TRGB catalog of EDD \citep{2009AJ....138..332J}. This procedure has been developed and matured with many years of work, and TRGB distances from the CMDs/TRGB catalog have served as key components for many results including the definition of our home supercluster Laniakea \citep{2014Natur.513...71T}, the realization of the effects of the neighbouring Local Void on the motions of nearby galaxies \citep{2008ApJ...676..184T,2017ApJ...835...78R,2019ApJ...880...52A}, and the determination of the extragalactic distance scale and the Hubble constant from the larger Cosmicflows programme \citep{2013AJ....146...86T,2016AJ....152...50T,2020ApJ...896....3K}. 

\subsection{Range of Distances}

\begin{figure*}
    \centering
    \includegraphics[width=\textwidth]{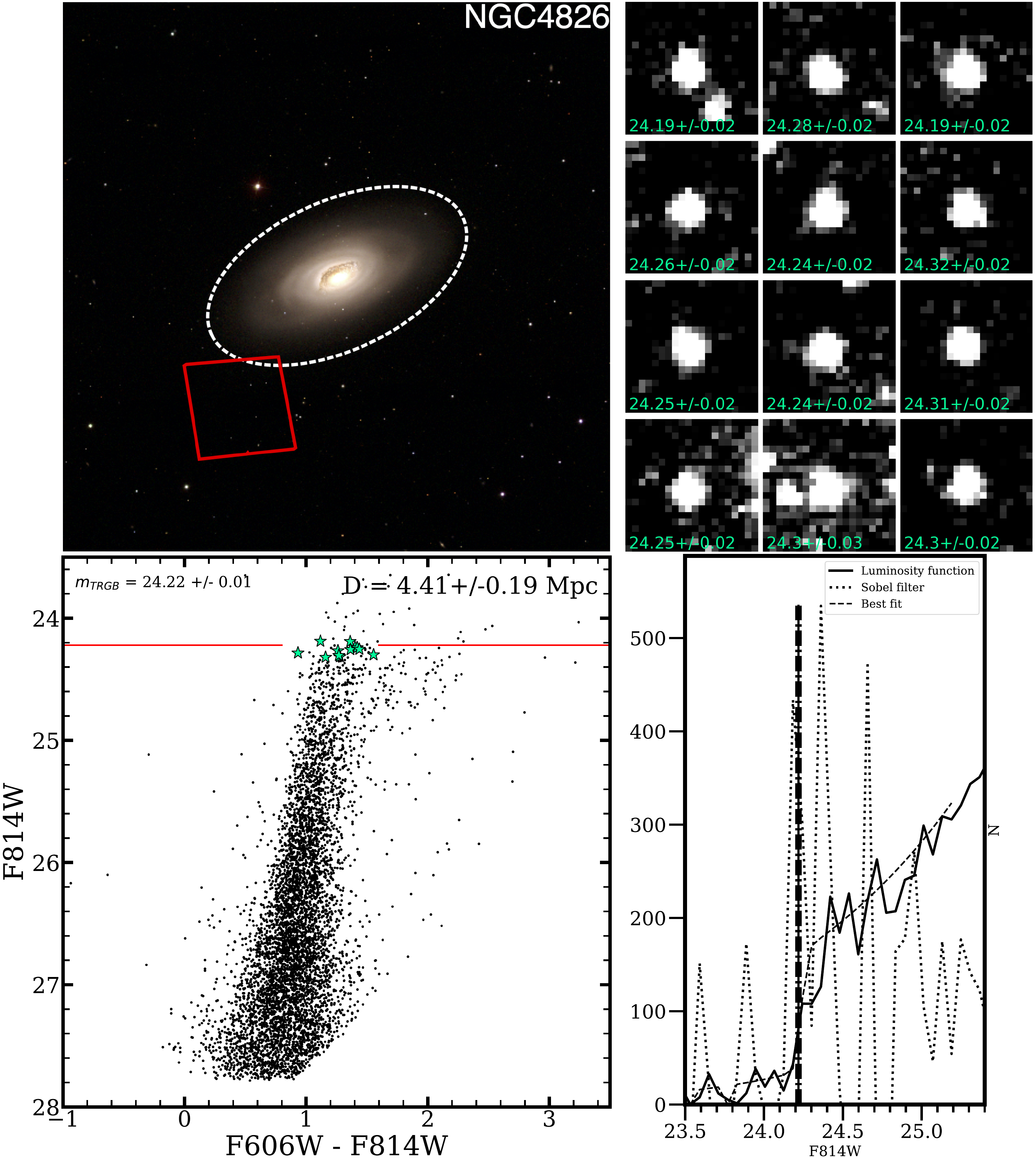}
    \caption{\textbf{Top Left:} PHANGS-HST parallel footprint (red) for NGC~4826 overlaid onto a \textit{gri} image of NGC~4826 from SDSS. $D_{\mathrm{25}}$ from RC3 \citep{1991rc3..book.....D} is shown in the dashed white lines. \textbf{Bottom Left:} CMD and TRGB determination (red line) from the portion of the field selected for analysis. The gap in the red line denotes the color range of stars used to measure the TRGB. \textbf{Top Right:} Cutouts of a sampling of 12 stars from within $\pm$0.1 mag of the measured TRGB, with measured DOLPHOT magnitudes and errors. These stars are highlighted as green stars on the CMD. \textbf{Bottom Right:} Luminosity function (solid line), Sobel filter measurement (dotted line, shown only for comparison), and best-fit luminosity function (dashed line, with errors shown as the vertical dash-dotted lines) from our analysis.}
    \label{fig:ngc4826}
\end{figure*}

\begin{figure*}
    \centering
    \includegraphics[width=\textwidth]{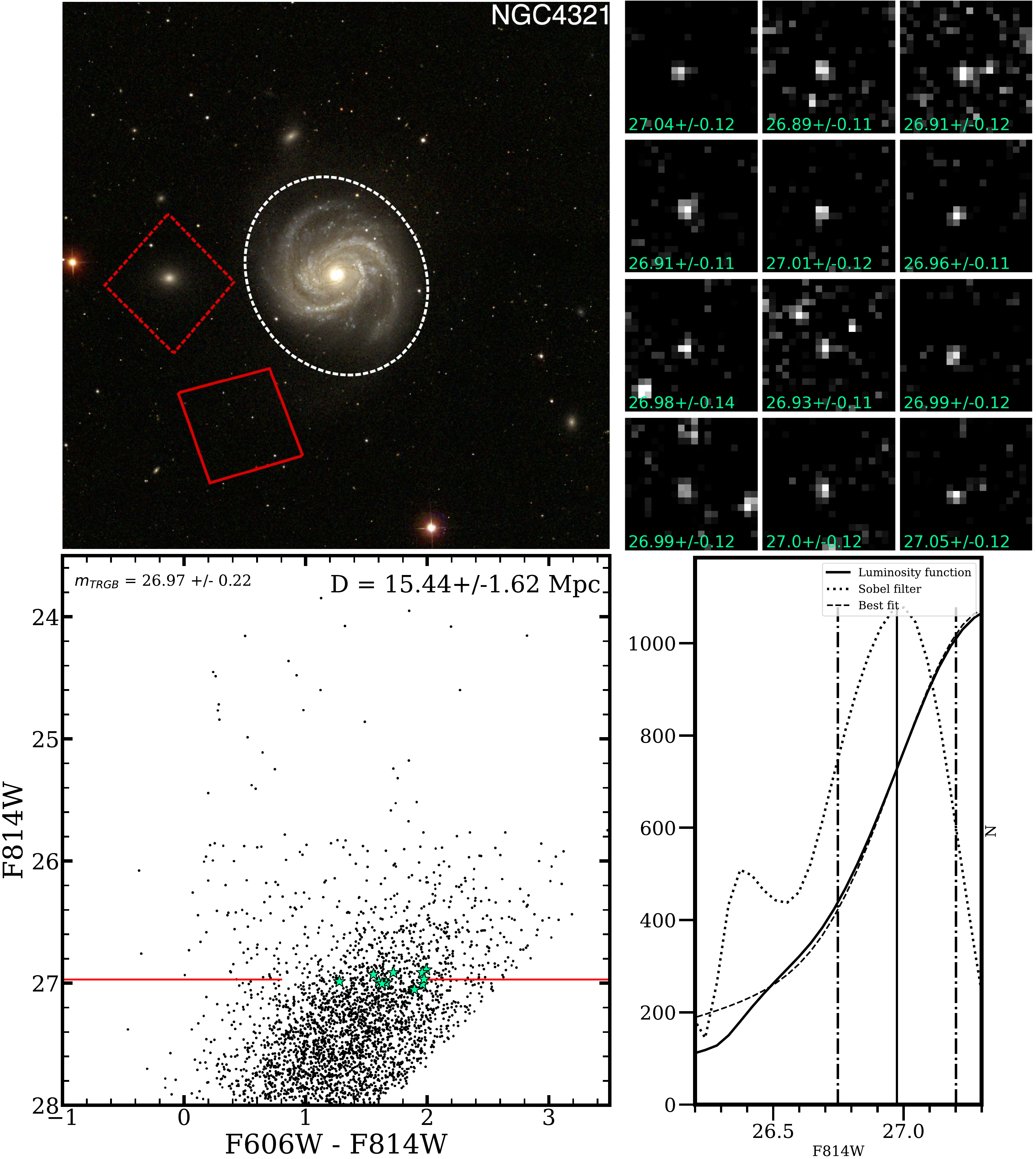}
    \caption{Same as Figure \ref{fig:ngc4826}, but for NGC~4321, which is at the far limit of our detection threshold for the TRGB with the PHANGS-HST parallel imaging. The parallel field highlighted with the dashed white square in the top-left panel falls on NGC~4328, a dwarf galaxy that is likely a satellite of NGC~4321. The analysis for this second parallel field is presented separately.}
    \label{fig:ngc4321}
\end{figure*}

In Figures \ref{fig:ngc4826} and \ref{fig:ngc4321}, we highlight the data for the closest (NGC~4826, $D$=4.4~Mpc) and furthest (NGC~4321, $D$=15.4~Mpc) PHANGS-HST targets for which we are successfully able to derive TRGB distances. In the top-left panels of these two figures, we show our \textit{HST} parallel imaging (red) overlaid on \textit{gri} footprints from the Sloan Digital Sky Survey (SDSS, \citealt{2000AJ....120.1579Y}), with $D_{\mathrm{25}}$ from RC3 \citep{1991rc3..book.....D} shown in dashed blue. Our main purpose in showing these figures is to draw attention to the quality of the data at these two extremes. At the near end, the stars are bright, well-resolved, and unambiguous. At the far end, we are approaching the limits of what can be achieved given the depth of the data, and the uncertainties involved become large.

In the top-right panels, we show F814W cutouts of a selection of a dozen stars within $\pm$0.1 mag of the measured TRGB, along with their DOLPHOT measured F814W magnitudes and errors. We note that as shown by previous studies \citep{2014ApJS..215....9W}, the error measurements from DOLPHOT are systematically underestimated. We emphasize that this does not affect our results. As previously discussed, the photometric errors that propagate into our final results are determined by the injection and recovery of artificial stars, which give proper estimates for photometric error.

The stars shown in the top-right panels are highlighted in green on the CMDs, along with the measured TRGB in the bottom-left panels. In the bottom-right panels, we show the observed luminosity function of stars, along with our best-fit from which we determine $m_{\mathrm{TRGB}}$ (with uncertainties indicated by the dotted-dashed lines). For purposes of comparison, we also provide the results of a Sobel filter (with a kernel of [$-$2, 0, 2]) on the same observed luminosity function. For the case of NGC~4826, the Sobel filter measurement is rather noisy (i.e., several peaks) due to the sparseness of the upper RGB, whereas our luminosity function fit is clean. For NGC~4321, both methods measure the same value of $m_{\mathrm{TRGB}}$ to within $\sim$0.02 mag.

\subsection{Measurements}
We now present the results of our TRGB analysis. Six of the galaxies for which we measure results already have existing TRGB distances in the literature. However, given the multi-orbit depth of the PHANGS parallel data, in four of these cases our measurements are improvements upon the existing work. In all of these cases, our new measurements agree very well with the existing literature measurements (typically within $\sim$2$\%$). We also present the first TRGB measurements for five galaxies, four of which represent the most precise distances to these galaxies to date. 

\subsubsection{Galaxies with Existing TRGB Measurements}

\begin{figure*}
    \centering
    \includegraphics[width=\textwidth]{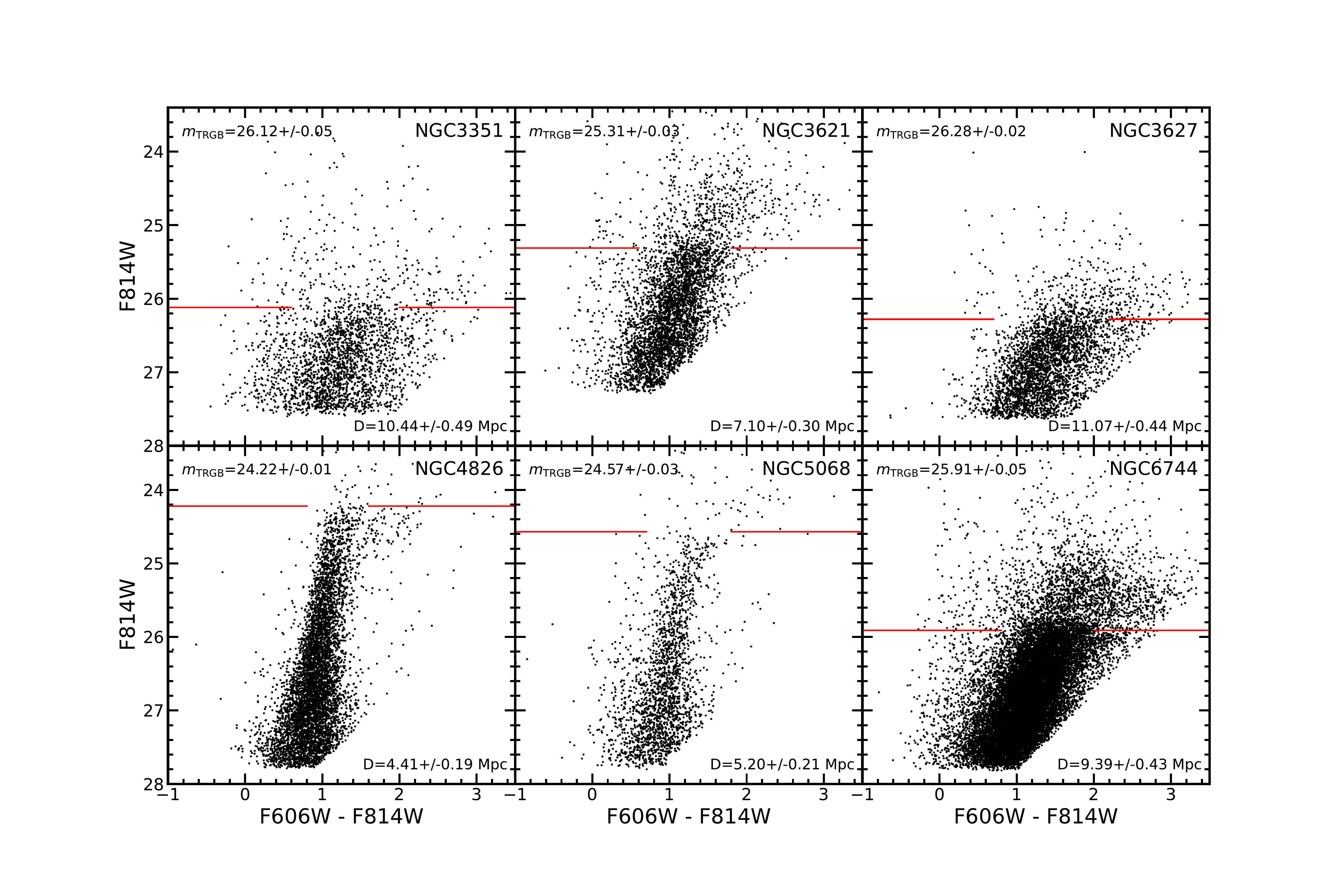}
    \caption{CMDs and TRGB measurements for the six PHANGS-HST galaxies with existing TRGB distances. The photometry (as plotted) has not been corrected for foreground reddening, and does not include the excluded regions with high levels of Population I stars. The gaps in the red lines denote the color ranges of stars used to measure the TRGB.}
    \label{fig:existingCMD}
\end{figure*}

\begin{figure*}
    \centering
    \includegraphics[width=\textwidth]{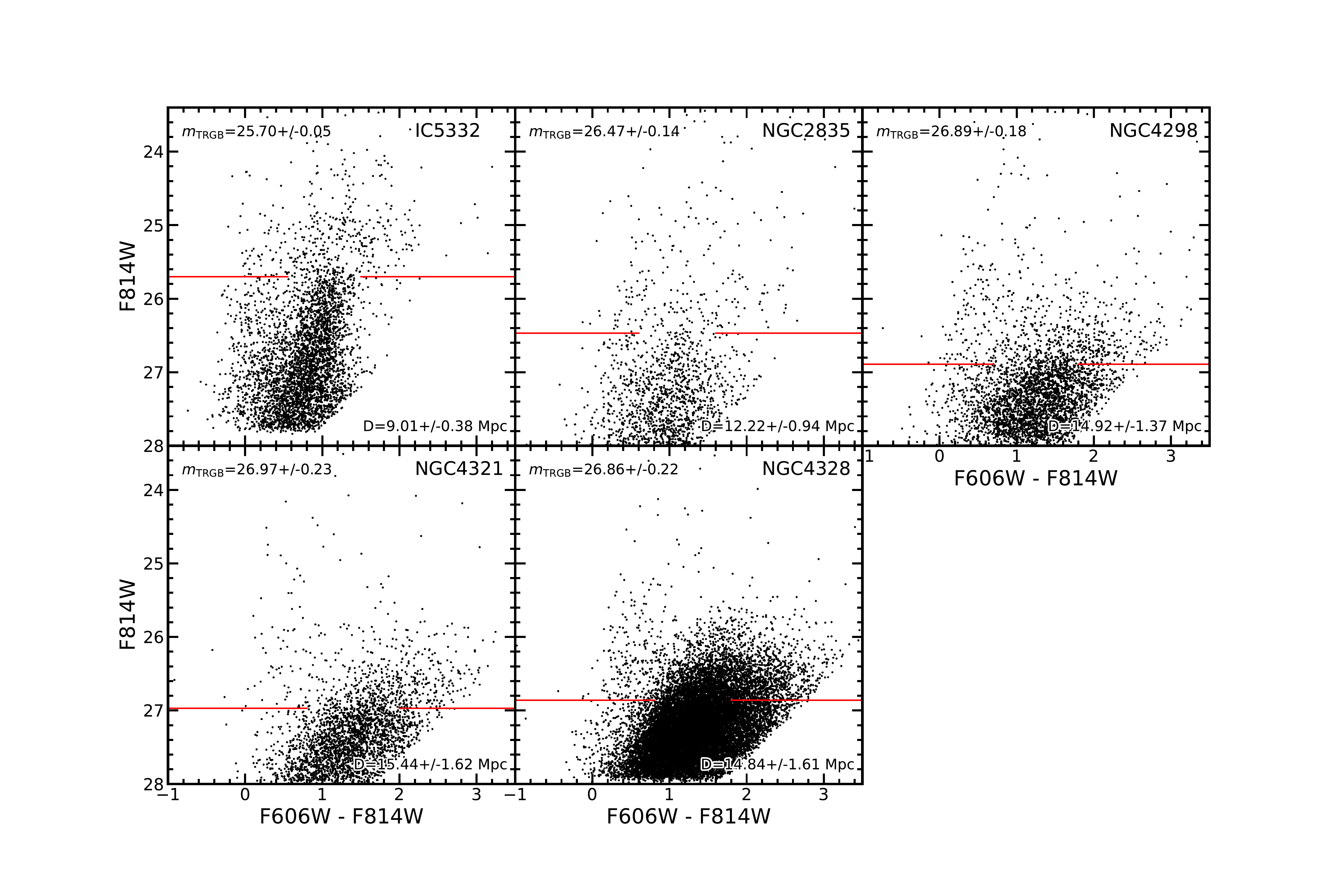}
    \caption{Same as Figure \ref{fig:existingCMD}, but for the five galaxies for which our TRGB measurements are the first.}
    \label{fig:newCMD}
\end{figure*}

\begin{itemize}
    \item NGC~3351 (M95) is a member of the Leo I group of galaxies, which includes the brighter NGC~3368 (M96) and NGC~3379 (M105). The CMDs/TRGB catalog's present TRGB measurement of $D$ = 9.96 $\pm$ 0.33~Mpc for NGC~3351 was obtained from older, \textit{HST} Wide Field and Planetary Camera 2 (WFPC2) observations of the galaxy's outer disk (GO-8584, PI: R.~Kennicutt). The new PHANGS-HST parallel data, whilst less contaminated with Pop I stars, has a sparser upper RGB---we measure $m_{\mathrm{TRGB}}$ = 26.12 $\pm$ 0.05 mag. The relative sparseness may inflate the measured distance to the target, which we find to be $D$ = 10.44 $\pm$ 0.49~Mpc. Due to the better population statistics in the archival data, we adopt the existing CMDs/TRGB catalog distance ($D$ = 9.96 $\pm$ 0.33~Mpc) to this galaxy, which is very close to the Cepheid determination of $D$ = 10.05 $\pm$ 0.42~Mpc \citep{2001ApJ...553...47F}.

    \item NGC~3621 is an isolated spiral galaxy and member of the Leo Spur, a filamentary structure whose members have, on average, relatively high negative peculiar velocities towards us. This observed effect is due to the expansion of the Local Void pushing galaxies residing in the Local Sheet (including the Milky Way) down towards the Leo Spur, which then imprints the negative peculiar velocities onto members of the Leo Spur \citep{2015ApJ...805..144K,2019ApJ...880...52A}. The existing CMDs/TRGB catalog measurement for NGC~3621 of 6.65 $\pm$ 0.18~Mpc was obtained from data taken by GO-9492 (PI: F.~Bresolin), with observations in the F555W and F814W bands. The usage of the F555W is less optimal, due to many of the higher metallicity (redder) RGB stars being pushed off to the right of the observable CMD. 
    
    PHANGS-HST provides two parallel fields, which are both deeper than other existing data. Both fields cover part of the outer disk, which we isolate to reduce contamination from Pop I stars. From the two fields (only the outermost one is shown in Figure \ref{fig:existingCMD}), we find  $m_{\mathrm{TRGB}}$ = 25.31 $\pm$ 0.03 and $m_{\mathrm{TRGB}}$ = 25.27 $\pm$ 0.02 mag, which result in $D$ = 7.10 $\pm$ 0.30~Mpc  and $D$ = 7.02 $\pm$ 0.28~Mpc, respectively. The results from these two fields agree very well, and we take an average of these two measurements with a conservative error estimate ($D = 7.06$ Mpc $\pm$ 0.28~Mpc) as the adopted distance to this galaxy.

    \item NGC~3627 is the brightest member of a group colloquially known as the Leo Triplet, though there are fifteen likely group members \citep{2017ApJ...843...16K}. There are several existing \textit{HST} observations that allow for the determination of a TRGB distance to this target, due to its nature as a host to SN1989B, a type Ia supernova. The most ideal data is from the Carnegie-Chicago Hubble Programme (CCHP, GO-13691, PI: W.~Freedman). The CCHP group, with their distinct methodology, has determined a distance to NGC~3627 of 11.06 $\pm$ 0.30 Mpc \citep{2019ApJ...882..150H, 2019ApJ...882...34F}. The existing determination on the CMDs/TRGB catalog from this same CCHP field is 11.32 $\pm$ 0.48~Mpc, implying the two measurements are consistent to within $\sim$2$\%$ of each other, though $\sim$8$\%$ offset from the Cepheid determination of \cite{2001ApJ...553...47F} ($D$ = 10.05 $\pm$ 0.37~Mpc).
    
    The PHANGS-HST field partially overlaps with the southern edge of the main disk, introducing some contamination from Pop I stars. We select the outer $\sim$30$\%$ of the observed field for our analysis. Within this region, we find $m_{\mathrm{TRGB}}$ = 26.28 $\pm$ 0.02 mag, from which we determine a distance of 11.07 $\pm$ 0.44~Mpc. We choose to adopt the existing CMDs/TRGB catalog distance over the new PHANGS-HST determination due to the greater number of detected stars in the selected regions of the field in the former, though we note that the difference between the adopted measurement and the PHANGS-HST determination is quite small ($\sim$2$\%$).
    
    \item NGC~4826, also known as the Black Eye galaxy due to its prominent dust lanes, is the closest PHANGS-HST target with a previously determined TRGB distance of $D$ = 4.40 $\pm$ 0.18~Mpc on the CMDs/TRGB catalog (from GO-10905, PI: R. Tully). The PHANGS-HST data samples a region of the galaxy where there are two distinct RGBs. This includes a low-metallicity population within the halo of the galaxy, as well as a high-metallicity population from the outer disk. There is evidence of a Type-III anti-truncation component \citep{2020ApJ...897..106K}, and this new parallel field will allow us to trace this component at further galactocentric radii (to be presented in a later work).
    
    For our TRGB analysis, we select the lower metallicity red giant population by limiting the analysis to the far half of the field. We find $m_{\mathrm{TRGB}}$ = 24.22 $\pm$ 0.01 mag. From this, we determine $D$ = 4.41 $\pm$ 0.19~Mpc, which is nearly identical to the existing determination on the CMDs/TRGB catalog. We note that our measurement of the TRGB is consistent within the small uncertainties with the recent determination of \cite{2020ApJ...897..106K}, who measure $m_{\mathrm{TRGB}}$ = 24.21 $\pm$ 0.03 mag from archival \textit{HST} data. We adopt the new PHANGS-HST TRGB determination for the distance to this galaxy.
    
    \item NGC~5068 is an isolated flocculent spiral, and the second-nearest PHANGS-HST target, with the existing CMDs/TRGB catalog measurement situating it at $D$ = 5.16 Mpc $\pm$ 0.19~Mpc. The PHANGS-HST field is nicely placed in the outer halo, from which we measure $m_{\mathrm{TRGB}}$ = 24.57 $\pm$ 0.03 mag. This equates to a distance of $D$ = 5.20~Mpc $\pm$ 0.21~Mpc. We choose to adopt the new PHANGS-HST measurement for the distance to NGC~5068, although the two measurements are within less than $1\%$ of each other.
    
    \item NGC~6744 is a Milky-Way like spiral and the brightest member of its group. 
    The CMDs/TRGB catalog measurement from data taken by SNAP-12546 (PI: R.~Tully) shows $D$ = 9.50 $\pm$ 0.63~Mpc. The PHANGS-HST parallel is deeper, and when limited to regions with fewer Pop I stars, provides a higher confidence measurement of $m_{\mathrm{TRGB}}$ = 25.91 $\pm$ 0.05 mag, resulting in a distance of $D$ = 9.39 $\pm$ 0.43~Mpc. We adopt the new PHANGS distance measurement for NGC~6744.
    
\end{itemize}

\subsubsection{Galaxies with First-Ever TRGB Measurements}

We now turn to galaxies which do not have prior TRGB measurements. Based on distance measures from other methods, these targets are on average further than those described in the previous subsection. 

\begin{figure*}
    \centering
    \includegraphics[width=\textwidth]{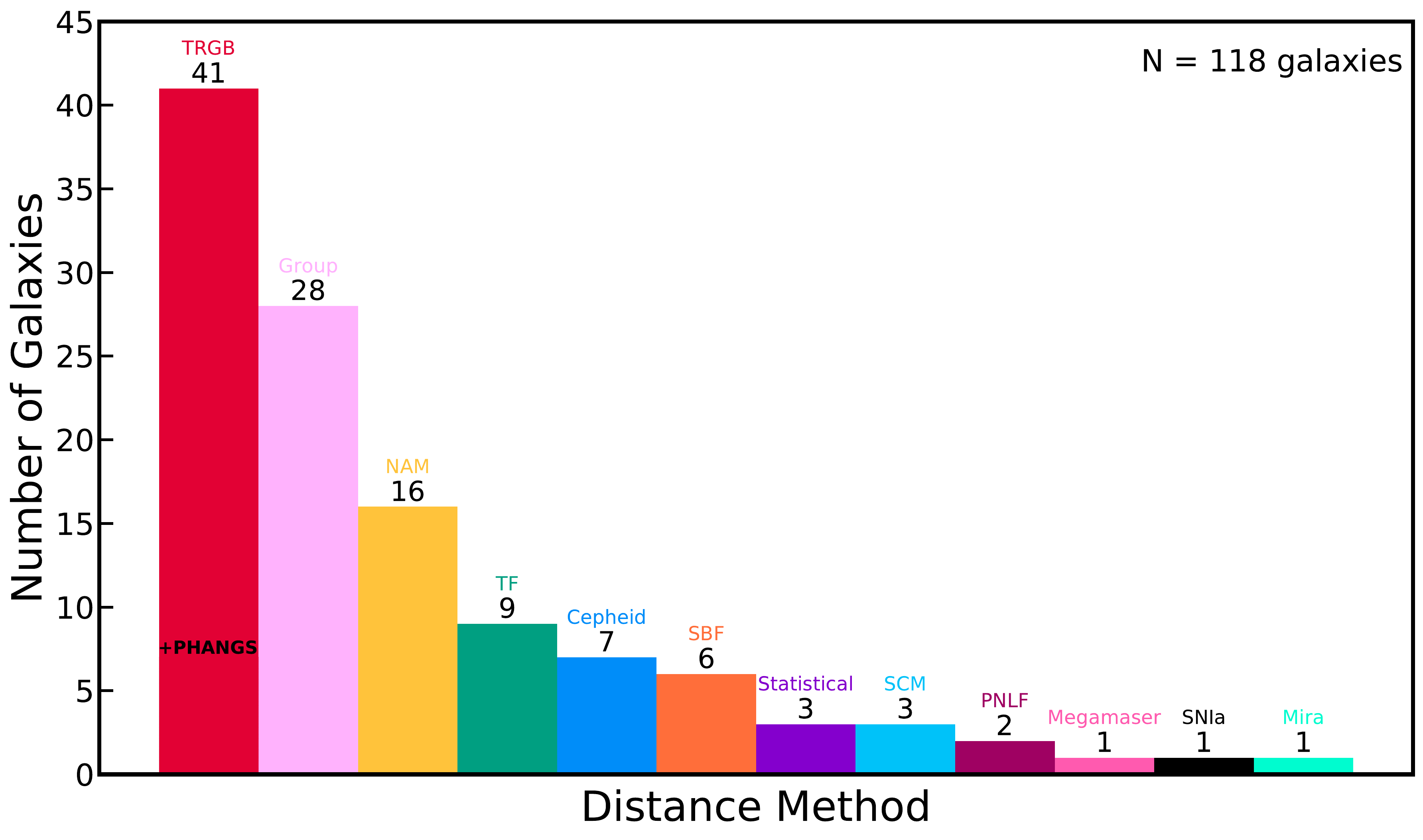}
    \caption{Histogram of adopted distances to PHANGS galaxies. New contributions from this paper are shown as the overlapping region on the TRGB bar.}
    \label{fig:distanceBar}
\end{figure*}

\begin{itemize}

    \item IC~5332 is a face-on spiral galaxy in the Sculptor constellation, but lying beyond the Sculptor group of galaxies. There was previously no reliable distance to this galaxy, with estimates relying on a tenuous group linkage with NGC~7713, or based on its recessional velocity. Our PHANGS-HST parallel is well-placed, with only a small portion of the field experiencing contamination from a young star cluster (which we remove). Based on this data, we measure $m_{\mathrm{TRGB}}$ = 25.70 $\pm$ 0.05 mag, which gives us $D$ = 9.01 $\pm$ 0.38~Mpc.

    \item NGC~2835 is the brightest member of a small galaxy group in the southern hemisphere. Most previous distances were based on the Tully-Fisher (TF) relation \citep{1977A&A....54..661T}, though with a nearly factor of 3 spread in the reported values. The numerical action method (NAM, \citealt{2017ApJ...850..207S}) distance (see Section 4 for more details) to this target is 12.38 $\pm$ 1.86~Mpc, which in the absence of a TRGB measurement represented the best distance estimate to this target. The PHANGS-HST parallel imaging for NGC~2835 lies far enough out into the halo of the galaxy that we do not need to trim the field for analysis due to the lack of a substantial population of young stars. From the entire field, we determine $m_{\mathrm{TRGB}}$ = 26.47 $\pm$ 0.14 mag, which results in our adopted distance of $D$ = 12.22 $\pm$ 0.94~Mpc. 
    
    \item NGC~4298 is a spiral galaxy that is a member of the Virgo Cluster. Previous best distance estimates relied on the TF relation, and measurements from Cosmicflows-3 found $D$ = 13.0 $\pm$ 3.0~Mpc. The PHANGS-HST parallel field lies in the combined halo of NGC~4298 and NGC~4302, a neighbouring edge-on spiral galaxy. The 21-cm HI maps for two galaxies show strong evidence for a bridge connecting the two galaxies \citep{2015ApJ...799...61Z}, though optical signatures of interaction are not obvious. We link the two galaxies together, and assume that they lie at the same distance. We use the stars in this parallel field to measure $m_{\mathrm{TRGB}}$ = 26.89 $\pm$ 0.18 mag, providing us with a new distance of $D$ = 14.92 $\pm$ 1.37~Mpc. 

    \item NGC~4321 (M100) is a large spiral galaxy located within the Virgo Cluster. Its location within Virgo, as well as the fact that it has been host to seven observed supernovae (including SN2006X, a type Ia) makes it an extremely important target for which to have an accurate distance. PHANGS-HST has two parallel fields for this target, one of which lies right on top of the nearby dwarf galaxy NGC~4328, which we discuss separately. The other parallel is relatively well-placed, and we use the outer portion of this field to determine $m_{\mathrm{TRGB}}$ = 26.97 $\pm$ 0.23 mag. From this, we derive $D$ = 15.44 $\pm$ 1.62~Mpc. Given the relatively large uncertainty in our distance, we adopt the Cepheid distance of $D$ = 15.21 $\pm$ 0.49~Mpc \citep{2001ApJ...553...47F} to this galaxy, though we note that our own determination is very close to this value.
    
    \item NGC~4328 is a dwarf galaxy that lies $\sim$6$'$ away from NGC~4321. Unlike NGC~4321's other satellite (NGC~4323), it is not clearly connected to NGC~4321, and thus could be physically unrelated. One of the two parallel fields for NGC~4321 fully covers this dwarf galaxy, and we are able to use the entire field (which lacks young stars) to measure $m_{\mathrm{TRGB}}$ = 26.86 $\pm$ 0.22 mag. From this we find $D$ = 14.84 $\pm$ 1.61~Mpc. With this distance, we confirm that NGC~4328 is likely a satellite of NGC~4321. Note that NGC~4328 is not a member of the PHANGS sample, and is included here only because one of the two parallels for NGC~4321 falls directly onto this satellite.

\end{itemize}

\subsubsection{Galaxies with Marginal/Null Results}
For the remainder of the PHANGS-HST sample, we are unable to determine a robust TRGB distance from the parallel data. The lack of results stems from the underlying data being too sparse (i.e., at too large a galactocentric radius), too shallow for the likely distance to the galaxy, or a combination of both. Here we briefly describe the marginal or null results from our data.

\begin{itemize}
    \item NGC~1317\footnote{Tied to measurements of its larger, interacting neighbour, NGC~1316.}, NGC~1365, and NGC~4536 all have precise measurements of the TRGB \citep{2009AJ....138..332J} obtained with much deeper \textit{HST} data (GO-13691, PI: W.~Freedman) than available from PHANGS-HST. Similarly, NGC~1559, NGC~4535, NGC~4548, and NGC~4654\footnote{Tied to measurements of its likely interacting neighbour, NGC~4639.} have precise distance measurements from Cepheid \citep{2001ApJ...553...47F} or Mira \citep{2020ApJ...889....5H} variables. These galaxies are all found to lie beyond $\sim$16~Mpc, which is beyond what is obtainable with the dataset presented in this paper.
    
    \item The remainder of the PHANGS-HST targets have too few resolved stars in their targeted fields, and/or do not reach a sufficient depth below the TRGB, and are thus not suitable for our purposes. These targets are NGC~1087, NGC~1097, NGC~1300, NGC~1385, NGC~1672, NGC~1792, NGC~2775, NGC~4254, NGC~4303, NGC~4569, NGC~4571, NGC~4689, and NGC~5248. Some of these galaxies (e.g., NGC~1087) have angular sizes small enough that their parallel fields simply lie too far out into the halo to be useful. For other targets (e.g., NGC~1300), the parallels fall at an appropriate galactocentric radius, but the galaxies are likely just too distant to detect enough RGB stars with the data.
\end{itemize}

\begin{figure*}
\centering
\includegraphics[width=\textwidth]{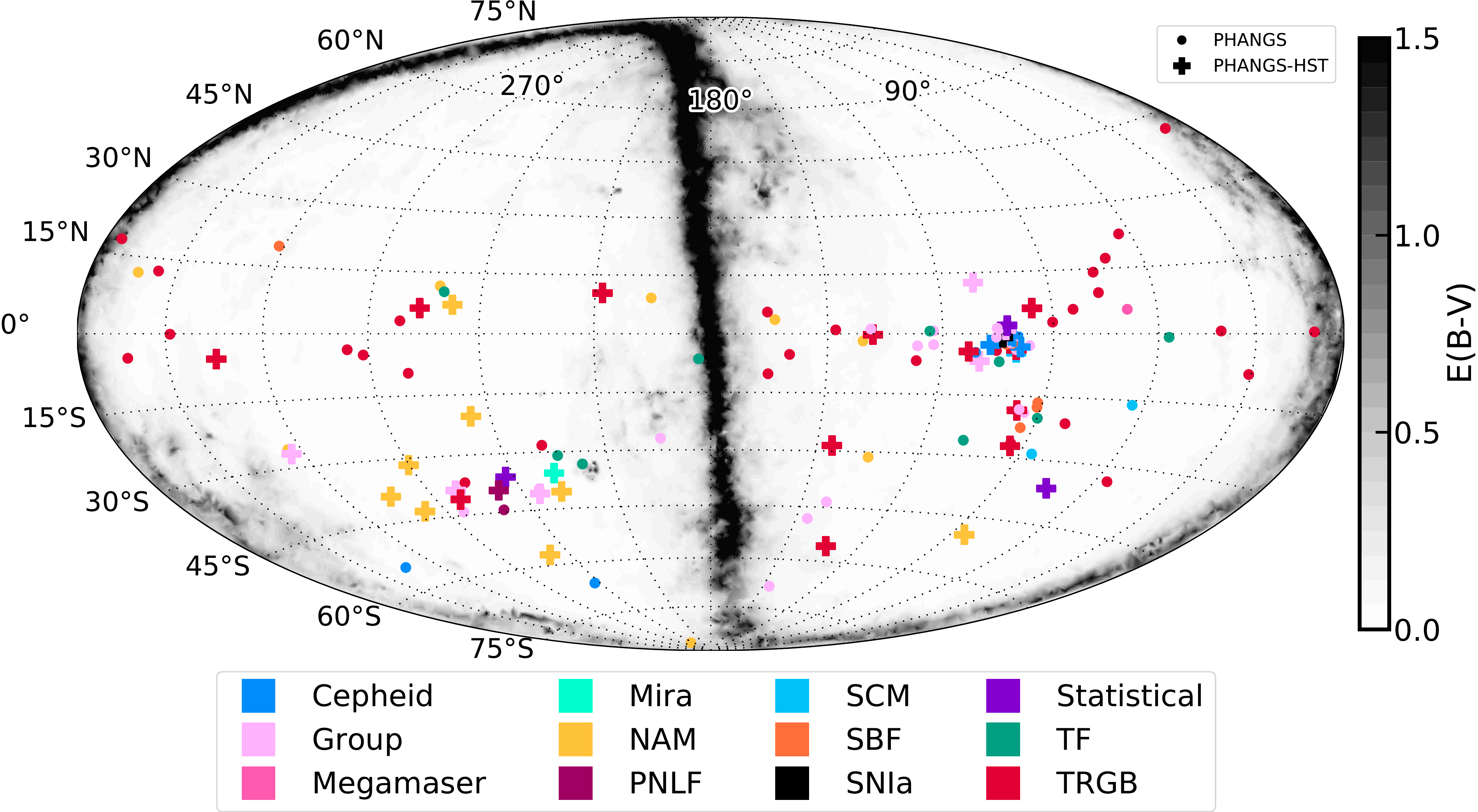}
\caption{A plot of the on-sky distribution of PHANGS galaxies, colour-coded by their adopted distance methods. We use the supergalactic coordinate system \citep{1991rc3..book.....D}. All of the 118 PHANGS targets of interest are shown, with those part of PHANGS-HST highlighted with a ``$+$" symbol. The foreground extinction is plotted to highlight the location of the Galactic plane, and is capped at a value of E(B-V) $=$ 1.5 for clarity.}
\label{fig:skyplot}
\end{figure*}


\section{Literature Distances}
\subsection{Distance Selections}
The PHANGS-HST galaxies make up a small but important subset (39/118) of the full PHANGS sample. To obtain distances to the the larger sample, as well as the remainder of the PHANGS-HST galaxies, we carefully combed through the available literature, a task aided by galaxy databases such as EDD and the NASA/IPAC Extragalactic Database (NED)\footnote{\url{http://ned.ipac.caltech.edu/}}. A summary figure with a histogram of our selected distances is presented in Figure \ref{fig:distanceBar}.\footnote{12-color colorblind friendly palette modified from \href{http://mkweb.bcgsc.ca/colorblind/}{``Designing for Color Blindness".}}

Due to the varying quality of individual measurements, we choose to not implement a strict hierarchy for the selection of distances. For instance, we rejected a couple of our TRGB measurements (NGC 1097 and NGC 1792) due to the possibility that they were actually measuring the onset of the AGB instead (see \cite{2019ApJ...872L...4A} for more details). In other cases, TF distances to galaxies with low inclinations (i.e., close to face-on) are often subject to large errors. Nearly every method of determining distances is subject to similar pitfalls, which necessitates a careful selection of the adopted distances.

We prioritise measurements from different distance techniques based on relative accuracy as has been demonstrated throughout the literature. Our first preference is for distances obtained from either the TRGB or Cepheid variables \citep{1912HarCi.173....1L}. Both of these techniques have been used extensively to find distances to nearby galaxies \citep{2001ApJ...553...47F, 2005MNRAS.356..979M,2009AJ....138..332J, 2016ApJ...826...56R,2019ApJ...882...34F}, and their overall accuracies (including systematic uncertainties) are $\sim$4--10$\%$ (based on quality of data, etc.). All of the archival TRGB measurements and errors are taken from the CMDs/TRGB Catalog of EDD \citep{2009AJ....138..332J}, with the errors standardised to include a 0.07 mag systematic error term added in quadrature to the measured statistical error \citep{2007ApJ...661..815R, 2017AJ....154...51M}. In this work, we have employed the same methodology as the TRGB measurements from the CMDs/TRGB catalog on EDD, hence minimizing internal systematic differences between the archival ($N$=33) and new PHANGS-HST ($N$=8) TRGB measurements. All but one \citep{1994Natur.371..385P} of the Cepheid variable measurements ($N$=7) are obtained from work done by the \textit{HST} Key Project \citep{2001ApJ...553...47F}---we choose to adopt their metallicity-corrected values and reported errors.

In the absence of highly accurate distances from either Cepheids or the TRGB, we turn to our next set of distance indicators. These include the standardizable candle method (SCM) for Type II supernovae \citep{2006ApJ...645..841N, 2015A&A...580L..15P}, surface brightness fluctuations (SBF,  \citealt{1988AJ.....96..807T}), the planetary nebula luminosity function (PNLF, \citealt{1989ApJ...339...53C, 1997ApJ...479..231F}), and the TF relation. As with the Cepheid and TRGB measurements, we aim to draw distances from larger, homogenised samples to minimise competing systematic errors. There are only 3 adopted distance measurements from the SCM, each from different sources. For these, we adopt the reported errors for each measurement. The PNLF distances ($N$=2) are obtained from MUSE observations of PHANGS galaxies (F.~Scheuermann et al., in preparation), and we adopt a preliminary error on each measurement of 10$\%$. All but one \citep{2004AJ....127.2031K} of the TF distances ($N$=9) are obtained from measurements from the Cosmicflows-3 programme \citep{2016AJ....152...50T}, and the SBF distances ($N$=6) are from a single large SBF programme \citep{2001ApJ...546..681T}, obtained through the Cosmicflows-3 catalog on EDD. For the TF and SBF distances, we adopt the error values as reported in Cosmicflows-3. 

\begin{figure*}
    \centering
    \includegraphics[width=\textwidth]{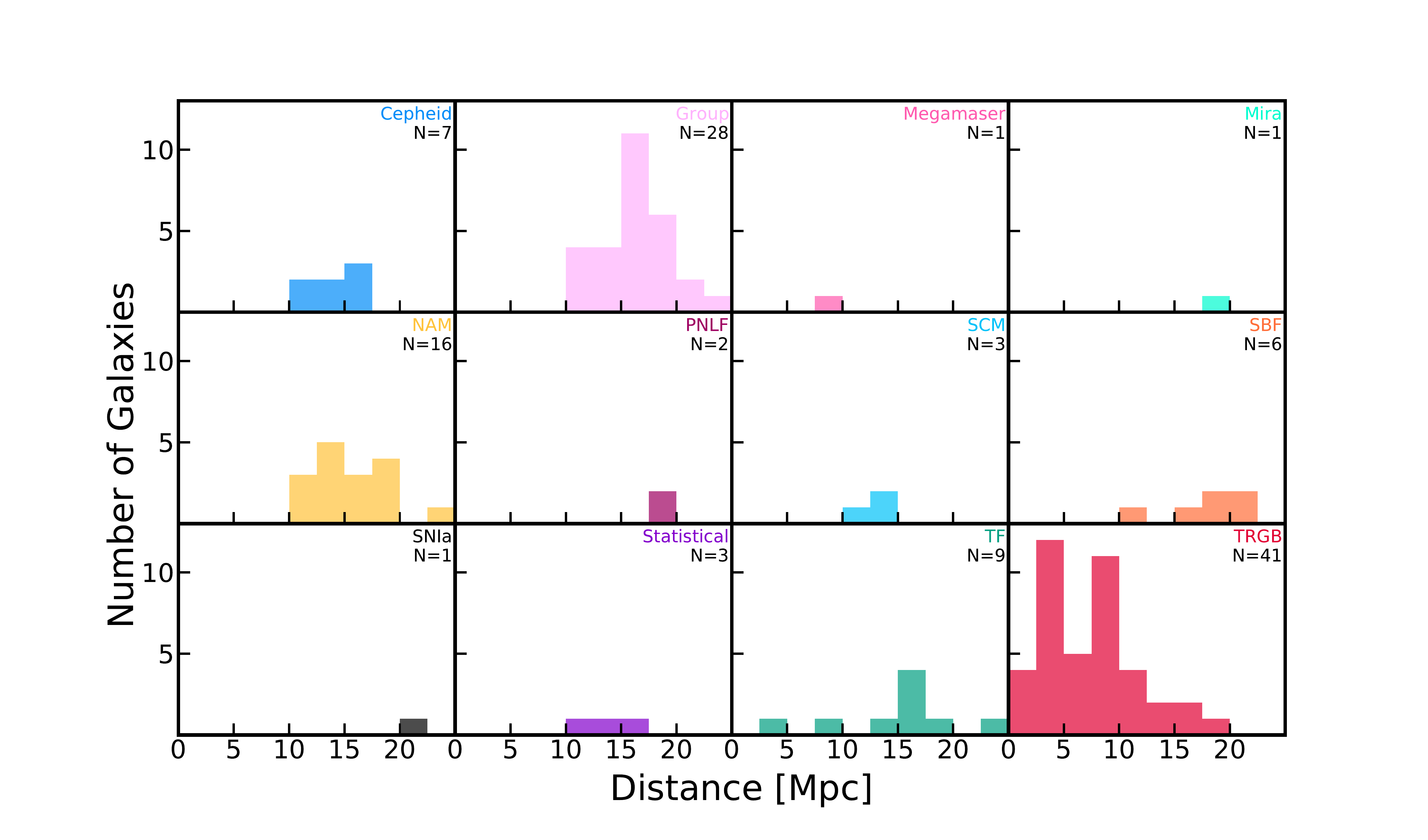}
    \caption{Histograms of each individual distance technique used in our compilation, highlighting the range of distances used for each method.}
    \label{fig:multiHist}
\end{figure*}
 
Finally, for galaxies without distance measurements from any of the above methods, we turn to distances from galaxy groups and numerical modeling of their orbits. For determining distances via galaxy groups ($N$=28), we use two different methods. The majority of group distances are obtained from the Kourkchi-Tully group catalog \citep{2017ApJ...843...16K}, which provides a robust catalog of galaxy groups within $\sim$3,500 km s$^{-1}$, or $D = 45$~Mpc. The catalog includes distances to groups, as well as errors on the \textit{group} distance measurement. To estimate our uncertainties for these group distances to \textit{individual} galaxies, we take the error in the group distance and multiply by the square root of the number of galaxies in the group with measured distances. This allows us to account for the varying physical sizes of different galaxy groups, since galaxies residing in physically larger groups will have larger uncertainties associated with their individual distances. In a few other cases, we tie PHANGS galaxy distances to individual galaxies with high-quality distances (e.g., from TRGB measurements). 
For instance, our adopted distance to NGC~1317 is a TRGB distance from a very deep \textit{HST} pointing \citep{2018ApJ...866..145H} of the halo of its likely interacting companion, NGC~1316. In these instances, we simply adopt the reported error on the original distance.

The last major distance indicator we use involves the usage of a numerical action methods (NAM) model \citep{2017ApJ...850..207S}. NAM is a non-linear model that attempts to reconstruct the 3-D orbits of galaxies from $z=4$ to the present day. The present iteration of NAM provides such information for nearly 1400 halos within 38 Mpc (which are embedded within a tidal field extending out to 100~Mpc).  \cite{2020AJ....159...67K} provides a smoothed velocity field derived from NAM, as well as an online distance-velocity calculator\footnote{\url{http://edd.ifa.hawaii.edu/NAMcalculator/}} to obtain either a distance or velocity given one of the two quantities, as well as a position on the sky. As much as possible, we avoid using NAM in heavily crowded galaxy environments due to the chaotic nature of the underlying velocity field and its poor correlation with distance. The most extreme example of this is in the Virgo Cluster, as galaxies in such a rich environment are heavily decoupled from the Hubble flow. For the galaxies with reported NAM distances ($N$=16), we report an uncertainty of 15$\%$ on the distance. For three galaxies where the TF measurements are less precise but similar to NAM, we choose to adopt a weighted average (referred to as ``statistical''). 

It is possible to directly compare galaxies with existing distance measurements from other methods, and the predicted value given by NAM. Figure 5 in \cite{2017ApJ...850..207S} shows a comparison of NAM distances and Hubble flow distances for 286 halos with high-quality distance measurements---NAM provides better distance estimates in most cases. However, given that these 286 distances served as inputs for NAM, an independent test of the reliability of NAM can only be performed in cases for which the distance to a galaxy was \textit{not} used as an input. Our new PHANGS-HST TRGB measurements provide two key new distances which can be used for this purpose. For instance, NAM predicts a distance to IC 5332 of 8.18 Mpc, with a nominal error of 15$\%$ ($\pm$1.23 Mpc). In this paper, we find the first accurate distance to this galaxy with a TRGB measurement of 9.01 $\pm$ 0.41 Mpc, or only $\sim$9$\%$ higher than the value predicted by NAM. In the case of NGC~2835, NAM predicts a value of 12.38 Mpc, whereas our measured TRGB distance is 12.22 $\pm$ 0.94 Mpc, a difference of only $\sim$1$\%$. These two cases, whilst limited, illustrate the predictive power of NAM to determine distances in cases where there are no other suitable measurements. 

In addition to the above methods, we note the selection of three additional distance measurement techniques, each of which was adopted for a single target. 
\begin{itemize}
    \item NGC~1559 has a recently published distance obtained from a newly derived period-luminosity relation for Mira variables \citep{2020ApJ...889....5H}. We adopt their distance of $D$ = 19.44 $\pm$ 0.44~Mpc.
    \item NGC~4258 is host to a water megamaser which allows the determination of a geometrical distance to the galaxy. We adopt the recent, highly-precise distance of $D$ = 7.58 $\pm$ 0.11~Mpc \citep{2019ApJ...886L..27R}.
    \item NGC~4579 is host to a type Ia supernova (SN 1989M), but has no corresponding Cepheid or TRGB distance. The lack of Cepheid distance is likely due to the fact that this is an older supernova with a somewhat poorly sampled light curve. Here we adopt a distance of $D$ = 21 $\pm$ 2~Mpc derived from observations of the SN 1989M \citep{1996ApJ...465L..83R}. 
\end{itemize}

\subsection{Compilation}
Our list of selected distances can be seen in Table~\ref{tab:distance-table}. Along with the PHANGS designation for the galaxy, we also provide the PGC number for each target \citep{2014A&A...570A..13M}, which allows for easier tracking between galaxy databases such as EDD and HyperLeda\footnote{http://leda.univ-lyon1.fr/}. For each galaxy, we specify whether it is one of the PHANGS-HST galaxies. Along with the adopted distance, we specify the distance error, distance method, and all references (original determination and any subsequent catalogs) from which the distance was obtained. For cases where the error is Gaussian on the distance modulus ($\mu$), we simply provide the larger error value as the reported error bar.

We show in Figure \ref{fig:skyplot} (inspired by figure 12 from \citealt{2017ApJ...843...16K}) the distribution of the entire PHANGS sample on the sky in supergalactic coordinates. Foreground extinction from \cite{2011ApJ...737..103S} is plotted in grey-scale, highlighting the location of the Galactic plane in these coordinates. Three different subsets of galaxies (archival ALMA, PHANGS-ALMA, or PHANGS-ALMA+HST) are denoted with different symbols, and galaxies are colour-coded by the final method used for selected distance. 
It can be seen from this plot that galaxies within this distance range are not uniformly distributed throughout the sky. The dearth of galaxies towards the north supergalactic pole corresponds to the location of the Local Void \citep{2018AJ....156..105A, 2019ApJ...880...24T}. Two main clusters of galaxies are also seen. One is loosely centered near $(250\degree ,-40\degree)$, and the other near $(100\degree ,-5\degree)$. These correspond to the Fornax and Virgo clusters, respectively. 

We show in Figure \ref{fig:distanceBar} a histogram of the adopted distances to galaxies in the PHANGS sample, and in Figure \ref{fig:multiHist} histograms for each individual distance method to highlight their distance distributions. A few general observations can be drawn from these figures and our sample:

\begin{itemize}
    \item The distance to nearly every PHANGS galaxy within 10~Mpc has been measured with the TRGB, highlighting our increasingly complete understanding of galaxy distances within the Local Volume. At present, most galaxies thought to lie within 10 Mpc without TRGB distances are small dwarfs, observations for many of which are currently being obtained and analyzed \citep{2020A&A...638A.111K} through a \textit{HST} Cycle 27 Snapshot programme (SNAP-15922, PI: R.~Tully). One key exception here is the Circinus Galaxy (ESO 097$-$G13), which is likely very nearby ($\sim$4~Mpc) and potentially crucial to the evolution of the Local Group \citep{2014MNRAS.440..405M, 2020MNRAS.494.2600N}. Unfortunatey, the galaxy is heavily obscured (the circle closest to the galactic plane in Figure \ref{fig:skyplot}) and would benefit from near-infrared observations with WFC3/IR to secure a robust TRGB distance.
    
    \item Distances obtained from observations of Cepheid variables as part of the Hubble Key Project \citep{2001ApJ...553...47F} are still the best distances for many galaxies at intermediate ($\sim$10--16~Mpc) distances, highlighting the long-lasting impact of this work. 
    
    \item With the increasing completeness of standard candle-based distances (e.g., Cepheid, TRGB) in the nearby universe, the reliance on less accurate techniques (e.g., SBF, TF) has decreased. Instead, these techniques are becoming increasingly popular \citep{2018ApJ...854L..31C,2020ApJ...896....3K} at much larger distances (out to $\sim$200~Mpc) and with much greater efficiency (many thousands of galaxies). This level of performance is simply not feasible for Cepheids/TRGB, which require much deeper, targeted observations with \textit{HST} (or future facilities).
    
    \item Distance estimates from NAM are valuable for many galaxies beyond 10 Mpc, especially since some of these targets lack \textit{any} velocity-independent distance. In instances for which there are only measurements from less precise methods (e.g., TF), NAM results provide an important reference point for cross-check.
    
\end{itemize}

\section{Summary}
We have successfully measured TRGB distances to 11 galaxies from the PHANGS-HST survey, from $\sim4$ to $\sim$15 Mpc, using imaging observations taken in parallel mode with ACS in the V and I bands (F606W, F814W). 
Five of these represent the first published TRGB distance measurements (Figure~\ref{fig:newCMD}), and eight are the best available distances to the targets (Table~\ref{tab:distance-table}).

Our analyses are based on the first year of observations through 2020 July, and include 37 ACS pointings in 30 galaxies.  Lack of TRGB measurement is due to the sparseness of the imaging (i.e., pointing at too large a galactocentric radius), insufficient depth of the 2-3 orbit observations given the likely distance to the galaxy, or a combination of both.  Results based on the remaining seven parallel fields (in six galaxies) will be presented in a short follow-up paper after the completion of the programme, which is anticipated in mid-2021. These parallel observations represent a valuable augmentation of the main PHANGS-HST science programme with no negative impact on the primary goals of the survey, and have provided a set of accurate distances without requiring a separate allocation of time on \textit{HST}. We recommend that future \textit{HST} programmes observing the disks of nearby galaxies to comparable or greater depths implement parallel halo observations for similar use.

In addition to the newly determined TRGB distances, we provide a compilation of the best available distances for the full sample of 118 PHANGS galaxies (Table~\ref{tab:distance-table}). These are the distances adopted by the first public PHANGS-ALMA data release (version 1.6). Updates will be made as improved distances become available, and will coincide with future public ALMA data releases (A.~K.\ Leroy et al., in preparation). 

We note that only about half of the PHANGS galaxies currently have distance measurements from highly reliable methods (e.g., Cepheids, TRGB), and that the majority of the remaining targets likely lie beyond 15~Mpc (Figure~\ref{fig:multiHist}). To obtain reliable distances to that large of a sample of galaxies with current facilities would require significant \textit{HST} time. For instance, accurate ($\sim$5\%) TRGB observations for galaxies at $\sim$20 Mpc require substantial time investments with \textit{HST}, such as the case for the type Ia supernova host NGC~1316 (16 orbits, \citealt{2018ApJ...866..145H}). It is unlikely that such expensive HST observations for the sole purpose of distance determination would be approved for \textit{every} PHANGS galaxy expected to lie at the far edge of our sample. 

Instead, future facilities such as the \textit{James Webb Space Telescope} (\textit{JWST}) will allow for much more efficient observations of the TRGB, due to a combination of its larger aperture and the brighter absolute magnitude of the TRGB in the near-infrared \citep{2014AJ....148....7W,2018SSRv..214..113B,2019ApJ...880...63M,2020ApJ...898...57D}. Similar strategies to the PHANGS-HST survey which employ observations of the disk in primary instrument and the halo in parallel can be applied to \textit{JWST}. Ideally, the role of WFC3 is replaced with the Mid-Infrared Instrument (MIRI) to obtain longer wavelength observations of the star-forming disk, and the job of observing halo stars would be accomplished with the Near Infrared Camera (NIRCam), instead of ACS.

\section*{Acknowledgements}
This work was carried out as part of the PHANGS collaboration. Based on observations made with the NASA/ESA Hubble Space Telescope, obtained from the data archive at the Space Telescope Science Institute. STScI is operated by the Association of Universities for Research in Astronomy, Inc. under NASA contract NAS 5-26555.  Support for Programme number 15654 was provided through a grant from the STScI under NASA contract NAS5-26555. We thank the anonymous referee for their useful comments which helped improve this manuscript.

G.A. acknowledges support from the IPAC Visiting Graduate Fellowship programme and from an award from the Space Telescope Science Institute in support of programme SNAP-15922. The work of AKL is partially supported by the National Science Foundation under Grants No. 1615105, 1615109, and 1653300. ES acknowledges funding from the European Research Council (ERC) under the European Union’s Horizon 2020 research and innovation programme (grant agreement No. 694343). K.K. and F.S. gratefully acknowledge funding from the German Research Foundation (DFG) in the form of an Emmy Noether Research Group (grant number KR4598/2-1, PI Kreckel). ER acknowledges the support of the Natural Sciences and Engineering Research Council of Canada (NSERC), funding reference number RGPIN-2017-03987. FS gratefully acknowledges funding from the German Research Foundation (DFG) Emmy Noether Research Group (grant number KR4598/2-1, PI Kreckel).  FB acknowledges funding from the European Union’s Horizon 2020 research and innovation programme (grant agreement No 726384/EMPIRE). MB acknowledges FONDECYT regular grant 1170618. RSK and SCOG acknowledge financial support from the German Research Foundation (DFG) via the Collaborative Research Center (SFB 881, Project-ID 138713538) 'The Milky Way System' (subprojects A1, B1, B2, and B8). He also thanks for funding from the Heidelberg Cluster of Excellence STRUCTURES in the framework of Germany's Excellence Strategy (grant EXC-2181/1 - 390900948) and for funding from the European Research Council via the ERC Synergy Grant ECOGAL (grant 855130). JMDK gratefully acknowledges funding from the Deutsche Forschungsgemeinschaft (DFG, German Research Foundation) through an Emmy Noether Research Group (grant number KR4801/1-1) and the DFG Sachbeihilfe (grant number KR4801/2-1), as well as from the European Research Council (ERC) under the European Union's Horizon 2020 research and innovation programme via the ERC Starting Grant MUSTANG (grant agreement number 714907). PSB acknowledges financial support from grant PID2019-107427GB-C31 from the Spanish Ministry of Economy and Competitiveness (MINECO). MQ acknowledges support from the research project PID2019-106027GA-C44
from the Spanish Ministerio de Ciencia e Innovaci\'on. TGW acknowledges funding from the European Research Council (ERC) under the European Union’s Horizon 2020 research and innovation programme (grant agreement No. 694343).

This research has made use of the NASA/IPAC Extragalactic Database (NED) which is operated by the Jet Propulsion Laboratory, California Institute of Technology, under contract with the National Aeronautics and Space Administration. This research has made use of the NASA/IPAC Infrared Science Archive, which is funded by the National Aeronautics and Space Administration and operated by the California Institute of Technology.

\section*{Data Availability}

The data underlying this article are available at the Mikulski Archive for Space Telescopes at \url{https://archive.stsci.edu/hst/search_retrieve.html} under proposal GO-15654. The photometry and list of derived parameters for the TRGB analysis are available under the CMDs/TRGB catalog of the Extragalactic Distance Database at \url{edd.ifa.hawaii.edu}.

\bibliographystyle{mnras}
\bibliography{paper} 

\newpage

$^{1}$IPAC, California Institute of Technology, Pasadena, CA 91125, USA\\
$^{2}$Institute for Astronomy, University of Hawaii, 2680 Woodlawn Drive, Honolulu, HI 96822, USA\\
$^{3}$Department of Astronomy, The Ohio State University,
4055 McPherson Laboratory, 140 West 18th Ave, Columbus, OH 43210, USA \\
$^{4}$Department of Physics, University of Alberta, 4-183 CCIS, Edmonton, Alberta, T6G 2E1, Canada \\
$^{5}$Max Planck Institut f\"{u}r Astronomie, Königstuhl 17, Heidelberg D-69117, Germany \\
$^{6}$Astronomisches Rechen-Institut, Zentrum f\"{u}r Astronomie der Universit\"{a}t Heidelberg, M\"{o}nchhofstra\ss e 12-14, 69120 Heidelberg, Germany \\
$^{7}$W.M. Keck Observatory, 65-1120 Mamalahoa Highway, Kamuela, HI 96743, USA \\
$^{8}$Department of Physics and Astronomy, The Johns Hopkins University, Baltimore, MD 21218, USA \\
$^{9}$Argelander-Institut f\"ur Astronomie, Universit\"at Bonn, Auf dem H\"ugel 71, 53121 Bonn, Germany \\
$^{10}$Observatories of the Carnegie Institution for Science, Pasadena, CA, USA \\
$^{11}$Departamento de Astronomía, Universidad de Chile, Santiago, Chile \\
$^{12}$Centro de Astronomía (CITEVA), Universidad de Antofagasta, Avenida Angamos 601, Antofagasta 1270300, Chile \\
$^{13}$Department of Physics $\&$ Astronomy, The University of Toledo, Toledo, OH 43606, USA \\
$^{14}$Department of Physics and Astronomy, University of Wyoming, Laramie, WY, USA \\
$^{15}$European Southern Observatory, Karl-Schwarzschild Straße 2, D-85748 Garching bei München, Germany \\
$^{16}$Univ Lyon, Univ Lyon, ENS de Lyon, CNRS, Centre de Recherche Astrophysique de Lyon UMR5574, F-69230 Saint-Genis-Laval, France \\
$^{17}$Universit\"{a}t Heidelberg, Zentrum f\"{u}r Astronomie, Institut f\"{u}r Theoretische Astrophysik,  Albert-Ueberle-Str. 2, 69120 Heidelberg, Germany \\
$^{18}$Research School of Astronomy and Astrophysics, Australian National University, Weston Creek, ACT 2611, Australia \\
$^{19}$International Centre for Radio Astronomy Research, The University of Western Australia, 35 Stirling Hwy, 6009 Crawley, WA, Australia \\
$^{20}$Universit\"{a}t Heidelberg, Interdisziplin\"{a}res Zentrum f\"{u}r Wissenschaftliches Rechnen, Im Neuenheimer Feld 205, 69120 Heidelberg, Germany \\
$^{21}$Astronomisches Rechen-Institut, Zentrum f\"{u}r Astronomie der Universit\"{a}t Heidelberg, M\"{o}nchhofstra\ss e 12-14, 69120 Heidelberg, Germany \\
$^{22}$Observatorio Astronómico Nacional (IGN), C/Alfonso XII 3, Madrid E-28014, Spain \\
$^{23}$Facultad de CC F\'{\i}sicas, Universidad Complutense de Madrid,  28040, Madrid, Spain \\
$^{24}$Max-Planck-Institut f\"ur Extraterrestrische Physik, Giessenbachstraße 1, D-85748 Garching bei München, Germany \\
$^{25}$Space Telescope Science Institute, 3700 San Martin Drive, Baltimore, MD 21218, USA


\appendix

\section{Exposure Times, Footprints, and Distances}

\begin{table}
\centering
\caption{Exposure times for all parallel fields from PHANGS-HST through 2020 July. Galaxies with two fields are denoted with F1 $\&$ F2 in order of increasing right ascension.}
\begin{tabular}{|l|c|c|}
\hline
\multicolumn{1}{|c|}{\textbf{Target}} & \multicolumn{1}{c|}{\textbf{F606W {[}s{]}}} & \multicolumn{1}{c|}{\textbf{F814W {[}s{]}}} \\ \hline
IC 5332                               & 3554                                        & 3215                                        \\ \hline
NGC~1087                              & 3536                                        & 3206                                        \\ \hline
NGC~1097 (F1)                         & 2051                                        & 2109                                        \\ \hline
NGC~1097 (F2)                         & 2051                                        & 2063                                        \\ \hline
NGC~1300 (F1)                         & 2017                                        & 2111                                        \\ \hline
NGC~1300 (F2)                         & 2053                                        & 2111                                        \\ \hline
NGC~1317                              & 3554                                        & 3215                                        \\ \hline
NGC~1365                              & 3556                                        & 3213                                        \\ \hline
NGC~1385                              & 3558                                        & 3217                                        \\ \hline
NGC~1559                              & 2070                                        & 2140                                        \\ \hline
NGC~1672 (F1)                         & 3063                                        & 3775                                        \\ \hline
NGC~1672 (F2)                         & 3063                                        & 3775                                        \\ \hline
NGC~1792                              & 3554                                        & 3215                                        \\ \hline
NGC~2775                              & 3544                                        & 3210                                        \\ \hline
NGC~2835                              & 3558                                        & 3217                                        \\ \hline
NGC~3351                              & 3554                                        & 3215                                        \\ \hline
NGC~3621 (F1)                         & 2051                                        & 2109                                        \\ \hline
NGC~3621 (F2)                         & 2051                                        & 2109                                        \\ \hline
NGC~3627                              & 3554                                        & 3215                                        \\ \hline
NGC~4254 (F1)                         & 3397                                        & 3348                                        \\ \hline
NGC~4254 (F2)                         & 3554                                        & 3215                                        \\ \hline
NGC~4298                              & 3419                                        & 3350                                        \\ \hline
NGC~4303                              & 3536                                        & 3206                                        \\ \hline
NGC~4321 (F1)                         & 3558                                        & 3217                                        \\ \hline
NGC~4321 (F2)                         & 3558                                        & 3217                                        \\ \hline
NGC~4535                              & 3554                                        & 3210                                        \\ \hline
NGC~4536 (F1)                         & 3536                                        & 3206                                        \\ \hline
NGC~4536 (F2)                         & 3531                                        & 3211                                        \\ \hline
NGC~4548                              & 3554                                        & 3215                                        \\ \hline
NGC~4569                              & 3554                                        & 3215                                        \\ \hline
NGC~4571                              & 3554                                        & 3215                                        \\ \hline
NGC~4654                              & 3554                                        & 3215                                        \\ \hline
NGC~4689                              & 3554                                        & 3215                                        \\ \hline
NGC~4826                              & 3558                                        & 3217                                        \\ \hline
NGC~5068                              & 3558                                        & 3217                                        \\ \hline
NGC~5248                              & 3554                                        & 3210                                        \\ \hline
NGC~6744                              & 3616                                        & 3246                                        \\ \hline
\end{tabular}
\label{tab:exposure-times}
\end{table}

\begin{figure*}
    \centering
    \includegraphics[width=\textwidth]{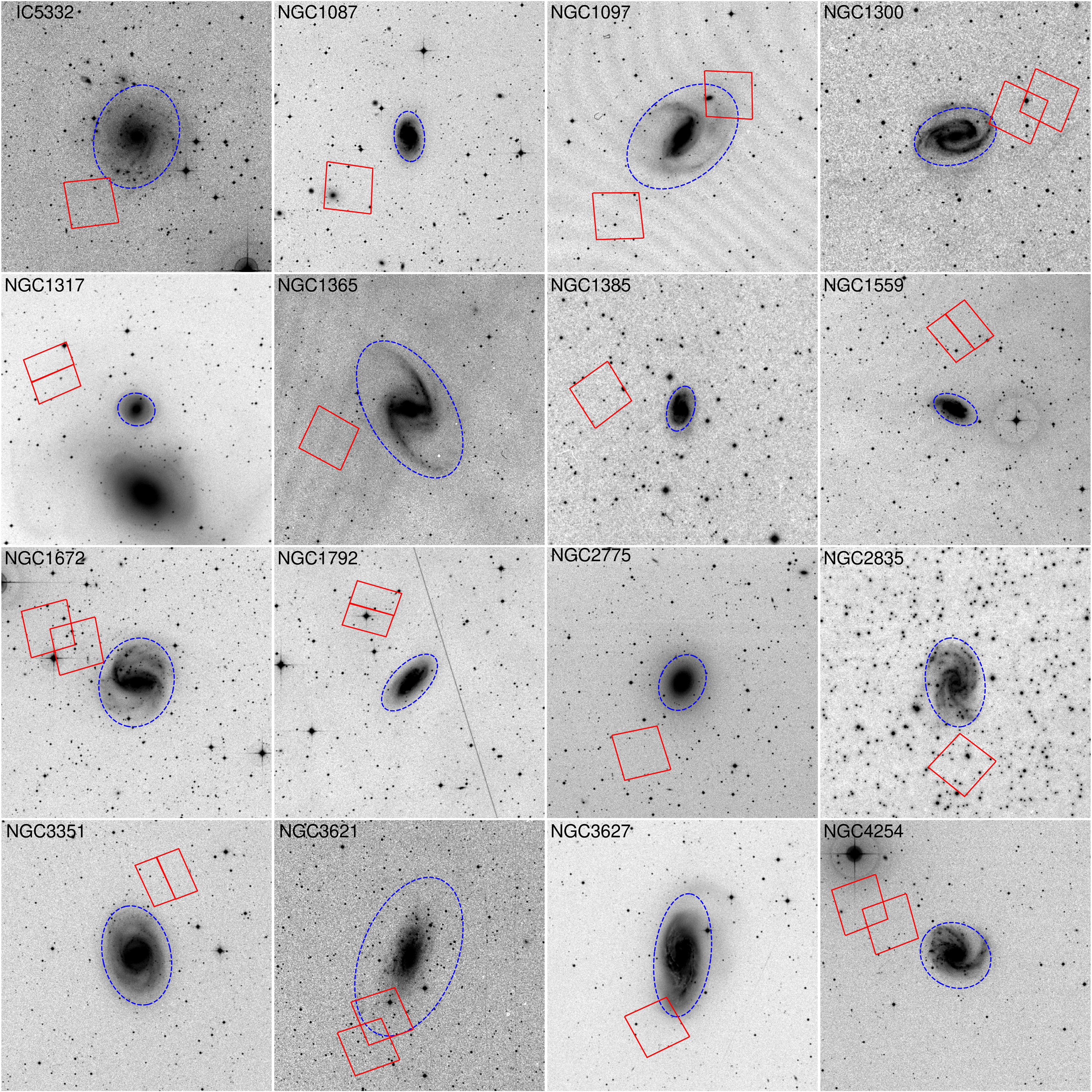}
    \caption{DSS footprints ($20\arcmin \times 20\arcmin$) for the first 16 galaxies observed in the first year of PHANGS-HST. The blue dashed lines indicate the location of $D_{25}$ from RC3, and the red regions show the locations of the ACS parallel imaging from PHANGS-HST.}
    \label{fig:fp1}
\end{figure*}

\begin{figure*}
    \centering
    \includegraphics[width=\textwidth]{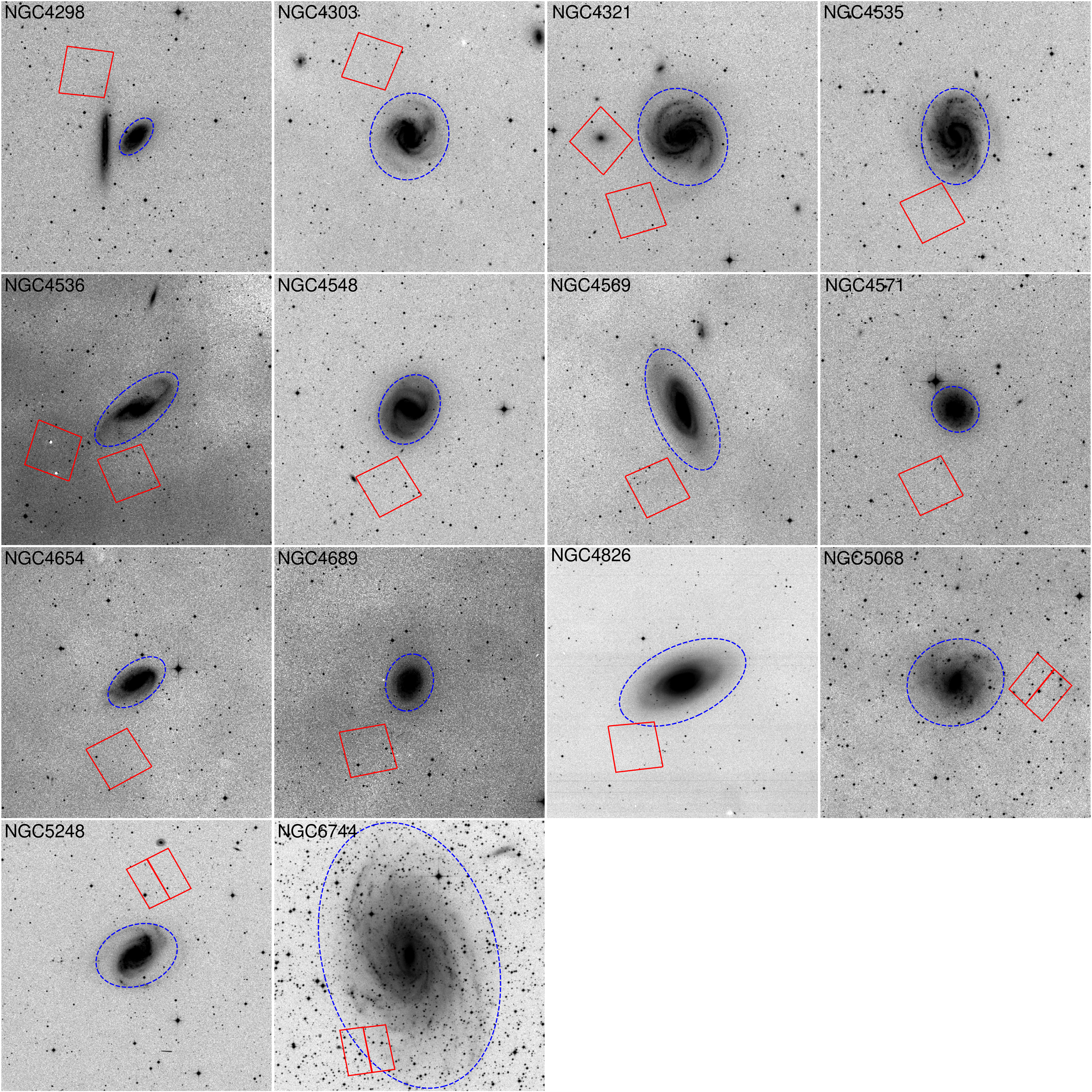}
    \caption{Same as figure \ref{fig:fp1}, for the remainder of the sample.}
    \label{fig:fp2}
\end{figure*}

\onecolumn
\centering

\onecolumn
\begin{longtable}[.85\textwidth]{lcccccc}
\caption{Distance compilation for the entire PHANGS galaxy sample ($N$=118). Galaxies with Y* in the PHANGS-HST column indicate that the \textit{HST} data for those is archival, and will be processed in the same way as the main PHANGS-HST sample. *We note that NGC~4328 is not part of the PHANGS sample, but is included in this table for completeness. \\ \textbf{References:} 
1) \citet{2004AJ....127.2031K}
2) \citet{2009AJ....138..332J} 
3) \citet{2016AJ....152...50T} 
4) \citet{2017ApJ...850..207S} 
5) \citet{2020AJ....159...67K} 
6) \citet{2017ApJ...843...16K} 
7) F. Scheuermann et al., in preparation
8) \citet{2020ApJ...889....5H} 
9) \citet{2003ApJ...594..247L} 
10) \citet{2001ApJ...553...47F} 
11) \citet{2010ApJ...715..833O}
12) \citet{2015MNRAS.448.2312B}
13) \citet{2001ApJ...546..681T}
14) \citet{2006ApJ...645..841N}
15) \citet{2019ApJ...886L..27R}
16) \citet{1994Natur.371..385P}
17) \citet{1996ApJ...465L..83R}} \\
\label{tab:distance-table}
\textbf{Galaxy} & \textbf{PGC Number} & \textbf{PHANGS-HST} & \textbf{Distance {[}Mpc{]}} & \textbf{Error {[}Mpc{]}} & \textbf{Method} & \textbf{Reference} \\ \hline
ESO~097-013      & 50779               &                     & 4.20                        & 0.77                     & TF              & 1               \\ \hline
IC~10            & 1305                &                     & 0.79                        & 0.05                     & TRGB            & 2               \\ \hline
IC~342           & 13826               &                     & 3.45                        & 0.13                     & TRGB            & 2               \\ \hline
IC~1954          & 13090               & Y                   & 12.8                       & 2.05                     & NAM+TF          & 3+4+5           \\ \hline
IC~1993          & 13840               &                     & 18.09                       & 2.71                     & Group           & 6               \\ \hline
IC~5273          & 70184               &                     & 14.18                       & 2.13                     & NAM             & 4+5             \\ \hline
IC~5332          & 71775               & Y                   & 9.01                        & 0.41                     & TRGB            & This Work       \\ \hline
NGC~224          & 2557                &                     & 0.82                        & 0.05                     & TRGB            & 2               \\ \hline
NGC~247          & 2758                &                     & 3.71                        & 0.13                     & TRGB            & 2               \\ \hline
NGC~253          & 2789                &                     & 3.70                        & 0.12                     & TRGB            & 2               \\ \hline
NGC~278          & 3051                &                     & 11.50                       & 1.73                     & NAM             & 4+5             \\ \hline
NGC~300          & 3238                &                     & 2.09                        & 0.09                     & TRGB            & 2               \\ \hline
NGC~598          & 5818                &                     & 0.94                        & 0.04                     & TRGB            & 2               \\ \hline
NGC~628          & 5974                & \phantom{a}Y*       & 9.84                        & 0.63                     & TRGB            & 2               \\ \hline
NGC~685          & 6581                & Y                   & 19.94                       & 2.99                     & NAM             & 4+5             \\ \hline
NGC~891          & 9031                &                     & 9.97                        & 0.45                     & TRGB            & 2               \\ \hline
NGC~1068         & 10266               &                     & 13.97                       & 2.10                     & NAM             & 4+5             \\ \hline
NGC~1087         & 10496               & Y                   & 15.85                       & 2.24                     & Group           & 6               \\ \hline
NGC~1097         & 10488               & Y                   & 13.58                       & 2.04                     & NAM             & 4+5             \\ \hline
NGC~1291         & 12209               &                     & 9.08                        & 0.52                     & TRGB            & 2               \\ \hline
NGC~1300         & 12412               & Y                   & 18.99                       & 2.85                     & NAM             & 4+5             \\ \hline
NGC~1313         & 12286               &                     & 4.32                        & 0.17                     & TRGB            & 2               \\ \hline
NGC~1317         & 12653               & Y                   & 19.11                       & 0.84                     & Group           & 2               \\ \hline
NGC~1326         & 12709               &                     & 18.34                       & 1.83                     & Group           & 6               \\ \hline
NGC~1365         & 13179               & Y                   & 19.57                       & 0.78                     & TRGB            & 2               \\ \hline
NGC~1385         & 13368               & Y                   & 17.22                       & 2.58                     & NAM             & 4+5             \\ \hline
NGC~1433         & 13586               & \phantom{a}Y*       & 18.63                       & 1.86                     & PNLF            & 7               \\ \hline
NGC~1511         & 14236               &                     & 15.28                       & 2.26                     & TF              & 3               \\ \hline
NGC~1512         & 14391               &                     & 18.83                       & 1.88                     & PNLF            & 7               \\ \hline
NGC~1546         & 14723               &                     & 17.69                       & 2.00                     & Group           & 6               \\ \hline
NGC~1559         & 14814               & Y                   & 19.44                       & 0.44                     & Mira            & 8               \\ \hline
NGC~1566         & 14897               & \phantom{a}Y*       & 17.69                       & 2.00                     & Group           & 6               \\ \hline
NGC~1637         & 15821               &                     & 11.70                       & 1.0                      & Cepheid         & 9               \\ \hline
NGC~1672         & 15941               & Y                   & 19.40                       & 2.91                     & NAM             & 4+5             \\ \hline
NGC~1792         & 16709               & Y                   & 16.20                       & 2.43                     & NAM             & 4+5             \\ \hline
NGC~1809         & 16599               &                     & 19.95                       & 5.63                     & TF              & 3               \\ \hline
NGC~2090         & 17819               &                     & 11.75                       & 0.84                     & Cepheid         & 10              \\ \hline
NGC~2283         & 19562               &                     & 13.68                       & 2.05                     & NAM             & 4+5             \\ \hline
NGC~2403         & 21396               &                     & 3.19                        & 0.13                     & TRGB            & 2               \\ \hline
NGC~2566         & 23303               &                     & 23.44                       & 3.52                     & Group           & 6               \\ \hline
NGC~2683         & 24930               &                     & 9.81                        & 0.43                     & TRGB            & 2               \\ \hline
NGC~2775         & 25861               & Y                   & 23.15                       & 3.47                     & NAM             & 4+5             \\ \hline
NGC~2835         & 26259               & Y                   & 12.22                       & 0.94                     & TRGB            & This Work       \\ \hline
NGC~2903         & 27077               & Y                   & 10.0                        & 2.5                      & NAM+TF          & 3+4+5           \\ \hline
NGC~2997         & 27978               &                     & 14.06                       & 2.81                     & Group           & 6               \\ \hline
NGC~3031         & 28630               &                     & 3.69                        & 0.21                     & TRGB            & 2               \\ \hline
NGC~3059         & 28298               &                     & 20.23                       & 4.05                     & Group           & 6               \\ \hline
NGC~3137         & 29530               &                     & 16.37                       & 2.32                     & Group           & 6               \\ \hline
NGC~3184         & 30087               &                     & 12.58                       & 1.74                     & SCM             & 11              \\ \hline
NGC~3239         & 30560               &                     & 10.86                       & 1.05                     & SCM             & 12              \\ \hline
NGC~3344         & 31968               &                     & 9.83                        & 1.27                     & TRGB            & 2               \\ \hline
NGC~3351         & 32007               & Y                   & 9.96                        & 0.33                     & TRGB            & 2               \\ \hline
NGC~3368         & 32192               &                     & 11.21                       & 0.49                     & TRGB            & 2               \\ \hline
NGC~3489         & 33160               &                     & 11.86                       & 1.62                     & SBF             & 3+13            \\ \hline
NGC~3507         & 33390               &                     & 23.55                       & 4.0                      & TF              & 3               \\ \hline
NGC~3511         & 33385               &                     & 13.94                       & 2.09                     & NAM             & 4+5             \\ \hline
NGC~3521         & 33550               &                     & 13.24                       & 1.97                     & TF              & 3               \\ \hline
NGC~3556         & 34030               &                     & 9.55                        & 1.41                     & TF              & 3               \\ \hline
NGC~3596         & 34298               &                     & 11.3                        & 1.1                      & Group           & 2               \\ \hline
NGC~3599         & 34326               &                     & 19.86                       & 2.73                     & SBF             & 3+13            \\ \hline
NGC~3621         & 34554               & Y                   & 7.06                        & 0.28                     & TRGB            & This Work       \\ \hline
NGC~3623         & 34612               &                     & 11.3                        & 1.1                      & Group           & 2               \\ \hline
NGC~3626         & 34684               &                     & 20.05                       & 2.34                     & SBF             & 3+13            \\ \hline
NGC~3627         & 34695               & Y                   & 11.32                       & 0.48                     & TRGB            & 2               \\ \hline
NGC~3628         & 34697               &                     & 11.3                        & 1.1                      & Group           & 2               \\ \hline
NGC~4207         & 39206               &                     & 15.78                       & 2.33                     & TF              & 3               \\ \hline
NGC~4254         & 39578               & Y                   & 13.1                        & 2.8                      & SCM             & 14              \\ \hline
NGC~4258         & 39600               &                     & 7.58                        & 0.11                     & Megamaser       & 15              \\ \hline
NGC~4293         & 39907               &                     & 15.76                       & 2.36                     & Group           & 6               \\ \hline
NGC~4298         & 39950               & Y                   & 14.92                       & 1.37                     & TRGB            & This Work       \\ \hline
NGC~4302         & 39974               &                     & 14.92                       & 1.37                     & Group           & This Work       \\ \hline
NGC~4303         & 40001               & Y                   & 16.99                       & 3.04                     & Group           & 6               \\ \hline
NGC~4321         & 40153               & Y                   & 15.21                       & 0.49                     & Cepheid         & 10              \\ \hline
NGC~4328*        & 40209               &                     & 14.84                       & 1.61                     & TRGB            & This Work       \\ \hline
NGC~4424         & 40809               &                     & 16.20                       & 0.70                     & TRGB            & 2               \\ \hline
NGC~4457         & 41101               &                     & 15.1                        & 2.3                      & Group           & 3+13            \\ \hline
NGC~4459         & 41104               &                     & 15.85                       & 2.18                     & SBF             & 3+13            \\ \hline
NGC~4476         & 41255               &                     & 17.54                       & 2.42                     & SBF             & 3+13            \\ \hline
NGC~4477         & 41260               &                     & 15.76                       & 2.36                     & Group           & 6               \\ \hline
NGC~4496A        & 41471               &                     & 14.86                       & 1.06                     & Cepheid         & 10              \\ \hline
NGC~4535         & 41812               & Y                   & 15.77                       & 0.37                     & Cepheid         & 10              \\ \hline
NGC~4536         & 41823               & Y                   & 16.25                       & 1.13                     & TRGB            & 2               \\ \hline
NGC~4540         & 41876               &                     & 15.76                       & 2.36                     & Group           & 6               \\ \hline
NGC~4548         & 41934               & Y                   & 16.22                       & 0.38                     & Cepheid         & 10              \\ \hline
NGC~4565         & 42038               &                     & 12.06                       & 0.43                     & TRGB            & 2               \\ \hline
NGC~4569         & 42089               & Y                   & 15.76                       & 2.36                     & Group           & 6               \\ \hline
NGC~4571         & 42100               & Y                   & 14.9                        & 1.2                      & Cepheid         & 16              \\ \hline
NGC~4579         & 42168               & \phantom{a}Y*       & 21.0                        & 2                        & SNIa            & 17              \\ \hline
NGC~4594         & 42407               &                     & 9.33                        & 0.47                     & TRGB            & 2               \\ \hline
NGC~4596         & 42401               &                     & 15.76                       & 2.36                     & Group           & 6               \\ \hline
NGC~4631         & 42637               &                     & 7.34                        & 0.27                     & TRGB            & 2               \\ \hline
NGC~4654         & 42857               & Y                   & 21.98                       & 1.16                     & Group           & 10              \\ \hline
NGC~4689         & 43186               & Y                   & 15.0                        & 2.25                     & NAM+TF          & 3+4+5           \\ \hline
NGC~4694         & 43241               &                     & 15.76                       & 2.36                     & Group           & 6               \\ \hline
NGC~4731         & 43507               &                     & 13.28                       & 2.12                     & Group           & 6               \\ \hline
NGC~4736         & 43495               &                     & 4.41                        & 0.16                     & TRGB            & 2               \\ \hline
NGC~4781         & 43902               &                     & 11.31                       & 1.18                     & Group           & 6               \\ \hline
NGC~4826         & 44182               & Y                   & 4.41                        & 0.19                     & TRGB            & This Work       \\ \hline
NGC~4941         & 45165               &                     & 15.0                        & 5.00                     & Group           & 6               \\ \hline
NGC~4945         & 45279               &                     & 3.47                        & 0.12                     & TRGB            & 2               \\ \hline
NGC~4951         & 45246               &                     & 15.0                        & 4.20                     & TF              & 3               \\ \hline
NGC~5042         & 46126               &                     & 16.78                       & 2.52                     & NAM             & 4+5             \\ \hline
NGC~5055         & 46153               &                     & 9.02                        & 0.33                     & TRGB            & 2               \\ \hline
NGC~5068         & 46400               & Y                   & 5.20                        & 0.21                     & TRGB            & This Work       \\ \hline
NGC~5128         & 46957               &                     & 3.69                        & 0.13                     & TRGB            & 2               \\ \hline
NGC~5134         & 46938               &                     & 19.92                       & 2.67                     & Group           & 6               \\ \hline
NGC~5194         & 47404               &                     & 8.56                        & 0.28                     & TRGB            & 2               \\ \hline
NGC~5236         & 48082               &                     & 4.89                        & 0.18                     & TRGB            & 2               \\ \hline
NGC~5248         & 48130               & Y                   & 14.87                       & 1.34                     & Group           & 6               \\ \hline
NGC~5457         & 50063               &                     & 6.65                        & 0.27                     & TRGB            & 2               \\ \hline
NGC~5530         & 51106               &                     & 12.27                       & 1.84                     & NAM             & 4+5             \\ \hline
NGC~5643         & 51969               &                     & 12.68                       & 0.53                     & TRGB            & 2               \\ \hline
NGC~6300         & 60001               &                     & 11.58                       & 1.74                     & NAM             & 4+5             \\ \hline
NGC~6744         & 62836               & Y                   & 9.39                        & 0.43                     & TRGB            & This Work       \\ \hline
NGC~6946         & 65001               &                     & 7.34                        & 0.68                     & TRGB            & 2               \\ \hline
NGC~7456         & 70304               &                     & 15.70                       & 2.33                     & TF              & 3               \\ \hline
NGC~7496         & 70588               & Y                   & 18.72                       & 2.81                     & NAM             & 4+5             \\ \hline
NGC~7743         & 72263               &                     & 20.32                       & 2.80                     & SBF             & 3+13            \\ \hline
NGC~7793         & 73049               &                      & 3.62                        & 0.15                     & TRGB            & 2               \\ \hline
\end{longtable}

\bsp	
\label{lastpage}
\end{document}